\documentclass[aps,pre,twocolumn,amsmath,amssymb,showpacs,amsfonts]{revtex4}
\usepackage{epsfig}
\usepackage{graphics}
\usepackage{dcolumn}
\usepackage{bm}
\usepackage{epstopdf}
\usepackage{color}
\usepackage{amsmath}

\begin{document}

\title{On phoretic clustering of particles in turbulence}
\author{Lukas Schmidt$^{1}$}
\author{Itzhak Fouxon$^{1, 2}$}
\author{Dominik Krug$^3$}
\author{Maarten van Reeuwijk$^{4}$}
\author{Markus Holzner$^1$}

\affiliation{$^1$ ETH Zurich, Stefano Franscini-Platz 5, 8093 Zurich, Switzerland}
\affiliation{$^2$ Department of Computational Science and Engineering, Yonsei University, Seoul 120-749, South Korea}
\affiliation{$^3$ Department of Mechanical Engineering, The University of Melbourne, Parkville, VIC 3010, Australia}
\affiliation{$^4$ Department of Civil and Environmental Engineering, Imperial College London, London SW7 2AZ, UK}

\begin{abstract}
We demonstrate that diffusiophoretic, thermophoretic and chemotactic phenomena in turbulence lead to clustering of particles on multi-fractal sets that can be described using one single framework, valid when the particle size is much smaller than the smallest length scale of turbulence $l_0$. To quantify the clustering, we derive positive pair correlations and fractal dimensions that hold for scales smaller than $l_0$. 
For scales larger than $l_0$ the pair correlation function is predicted to show a stretched exponential decay towards 1. In the case of inhomogeneous turbulence we find that the fractal dimension depends on the direction of inhomogeneity.
By performing experiments with particles in a turbulent gravity current we demonstrate clustering induced by salinity gradients in conformity to the theory.
The particle size in the experiment is comparable to $l_0$, outside the strict validity region of the theory, suggesting that the theoretical predictions transfer to this practically relevant regime.
This clustering mechanism may provide the key to the understanding of a multitude of processes such as formation of marine snow in the ocean and population dynamics of chemotactic bacteria.
\end{abstract}

\maketitle

\section{Introduction}

Inhomogeneous random distributions of advected fields like temperature, concentrations of salt or nutrients occur ubiquitously in fluids due to turbulence \cite{Frisch,tennekes}. For particles that perform phoresis (i.e.\ steady drift) in the gradients of the convected fields, the fields' inhomogeneities imply a finite velocity difference between the local flow and the particles \cite{and}. Particles that perform thermophoresis in a fluid at rest (steady drift in constant temperature gradient) will drift through thermal convection flow and particles that perform diffusiophoresis (steady drift in constant gradient of salinity) will drift through the turbulent ocean. Thus while the turbulence is incompressible so that the steady-state distribution of tracers is uniform, the distribution of particles that perform phoresis can be inhomogeneous. This holds independent of the flow regime. Volk et al.\cite{volk} were among the first to describe this phenomenon in a non-laminar flow environment by performing simulations in the context of chaotic flows. In this work we focus entirely on turbulent flows, we construct a quantitative theory of clustering of phoretic particles in turbulence and then demonstrate diffusiophoretic clustering experimentally in the range of parameters inaccessible by the theory. This is followed by the conclusion that the particle distribution occurs on a multifractal set with power-law pair correlations. If not stated otherwise particles considered throughout the entire study are small, light and spherical particles whose velocity relaxation time is much smaller than the Kolmogorov time of turbulence \cite{Frisch}.

Preferential concentration is well-studied in the case of inertial particles \cite{Maxey,FFB,FFS,review,Bec,Falkovich,BGH,Collins,BecCenciniHillerbranddelta,Stefano,Cencini,Olla,MehligWilkinson,MW,Shaw,fouxon1,FP1,Bewley,caustics,BecRaf,FouxonHorvai,Bec2006,JYL,Simonin,Kaufmann,Fevr,Reeks,rum,Bala,Elg} where it plays an important role in a wide range of phenomena including aerosols spreading in the atmosphere \cite{Seinfeld, Flagan}, planetary physics \cite{planetary1}, transport of materials by air or by liquids \cite{Engineering1}, liquid fuel combustion engines \cite{Engineering2}, rain formation in liquid clouds \cite{FFS,review,fouxon1,fphl} and many more.
Inertia, in the case of small particles, produces a small but finite difference between the particle's velocity and the local velocity of the fluid.
This difference is determined uniquely by the local flow so the particles' motion in space is a smooth flow given by the local turbulent flow corrected by drift.
Despite the smallness of the drift component, it results in a compressible particle flow causing accumulation of particles with time in preferred regions of the flow and an inhomogeneous steady state distribution \cite{Maxey}; this parallels the ordinary centrifuge where uniform initial distributions of inertial and tracer particles become completely different with time. While inertial particles, however small their inertia is, will eventually accumulate on the boundary of the centrifuge, the tracer distribution will always stay uniformly distributed.
In the case of turbulence the "boundary" on which inertial particles concentrate becomes very complex and time-dependent but it still has zero volume being multi-fractal \cite{FFS,Bec,fouxon1}.

The statistics of preferential particle concentration in turbulent flows obeying the incompressible Navier-Stokes equations can be described theoretically in the universal framework of weakly compressible flows \cite{fouxon1,FFS,fphl}, for not too heavy particles much smaller than the Kolmogorov lengthscale.
The complete statistics of the particle concentration where fluctuations are non-negligible (small-scale turbulence) depend on the statistics of turbulence through a single parameter $\Delta$, which provides the scaling exponent of the power-law correlations of the particle concentration.
Outside the viscous range of small-scale turbulence the fluctuations of the particle concentration are small. Thus different flows (large or smaller Reynolds number, including chaotic spatially uniform random flows) characterized by identical values of $\Delta$ will have identical statistics of the transported particle distribution.

These universal statistics are the reason for recent observations of preferential concentration of living phytoplankton cells \cite{Nature2013}.
Though single cell motion is very different from the one of an inertial particle, both can be described with smooth spatial flow in a range of parameters.
The flows are quite different but both are weakly compressible. Theory then implies identical statistics of inertial particles and phytoplankton which is confirmed experimentally.

Several studies provide indirect evidence that preferential concentration can be induced by phoresis as well. Diffusiophoretic drift (due to salinity gradients) has been observed in microfluidic laminar flows \cite{abecassis}, and has been shown to significantly affect the particle distributions. Recent experimental and numerical investigations have provided additional insight into the effect of diffusiophoresis in chaotic flows \cite{deseigne,volk}.
Furthermore, numerical simulations have shown that chemotactic bacteria may accumulate in nutrient patches in a turbulent flow \cite{TaylorStocker}. Thermophoresis leads to increased particle concentration in temperature minima or away from minima depending on their inertia \cite{Belan}.
However, clustering due to any kind of phoresis in fully turbulent flows has neither been observed nor described theoretically.

In this paper we extend the universal framework for weakly compressible flow \cite{fouxon1} to phoretic particles, leading to a prediction for the fractal dimension of the expected particle concentration in an inhomogeneous turbulent flow.
Our theoretical considerations are constructed for particles with comparatively small size and small velocity relaxation time that parallels the regime of particles with small but non-negligible inertia.
The theory is validated experimentally using high-frequency 3-D velocity and density measurements of diffusiophoretic particles in a fully turbulent gravity current.\\
The paper is structured as follows. In Section \ref{sec:phoresis}, a general introduction of phoretic particles and their governing equations is provided. The Sections \ref{sec:microscopic}\,\&\,\ref{sec:fluidmech} describe microscopic and macroscopic frameworks for the description of phoretic phenomena in macroscopically moving fluid, Section \ref{sec:smallscale} introduces relevant properties of small-scale turbulence. The theory of clustering of small particles in homogeneous turbulence is described in Section \ref{section:2}, and is extended to inhomogeneous turbulence in Section \ref{derivation}.
The theoretical study of pair correlations outside the scale of smoothness is provided in Section \ref{sec:five}.
The results from laboratory experiments of a turbulent gravity current are discussed in Section \ref{section4}, and demonstrate phoretic clustering in agreement with the theory, despite having particle sizes comparable to the Batchelor scale which are  formally outside of the validity region of the theory.
Concluding remarks are made in Section \ref{sec:conclusions}, including the implications these findings may have on the  formation of marine snow, the settling of organic particle aggregates in the ocean serving as deep-sea nutrient supply.

\section{Phoresis in turbulent flows}
\label{sec:phoresis}

Phoresis is a universal phenomenon of steady drift of macroscopic particles in an inhomogeneous motionless medium due to gradients in a scalar field $\phi(\bm x)$.
Gradients in $\phi$ cause a difference in forcing on different sides of the particle's surface, resulting in particle motion.
In probably the simplest instance of this phenomenon - thermophoresis in gases - the force is caused by a difference of the intensity of collisions with particles of inhomogeneously heated gas. The scalar field $\phi$ is temperature in this case. The unbalance in collisions causes particle drift toward the colder regions of the fluid.

In the general case the direction of movement depends on the underlying physics of the phoresis. However, when isotropy holds, the motion is parallel to the gradient of the field so the phoretic drift velocity $\bm v_{ph}$ is generally of the form
\begin{eqnarray}&&
\bm v_{ph}=c_{ph}\nabla\phi,\label{phorbs}
\end{eqnarray}
where $c_{ph}\left(\bm x(t), \phi[\bm x(t)]\right)$ is the phoretic coefficient that can depend on the particle position $\bm x(t)$ through the dependence on $\phi$ or other local fields (e.g density). It is assumed that the variation of $\phi(\bm x)$ over the particle's size is small. If it is not then higher powers of $\nabla\phi$ and higher-order derivatives of $\phi$ contribute to $\bm v_{ph}$. The phoretic velocity $\bm v_{ph}$ is attained after transients that take a finite relaxation time $\tau_{rel}$ during which the particle passes a characteristic distance $v_{ph}\tau_{rel}$.

The phoretic coefficient $c_{ph}$ can be both positive or negative, and will depend on the type of phoresis, cf. Table \ref{table}.
In the case of thermophoresis the particle reacts to the gradient of temperature of the fluid $T$ so that $\phi=T$. In gases the random hits of macroscopic particle from the gas molecules are stronger at the particles' side closer to higher temperature fluid. The particle is driven to regions with lower temperature so $c_T$ in Table \ref{table} is negative. In liquids or in gases when small particles are considered, the interactions are more complex and both signs of $c_T$ can hold, see \cite{pr,hottovy} and references therein.

Diffusiophoresis is the drift of a colloidal particle in response to a gradient of the concentration $C$ of a molecular solute \cite{andersonprieve,and}. For electrolyte solutions (such as saltwater), which will be studied experimentally in Section \ref{section4}, the drift velocity obeys $\bm v_{ph}\,\approx\,D_p \nabla\ln C$. Where $D_p$ is the diffusiophoretic constant that describes electrical and chemical couplings in the interfacial region between the particle surface and the surrounding solute inducing the drift \cite{andersonprieve,abecassis}. The diffusiophoretic constant depends on the particle's zeta $(\zeta)$-potential (a measure for the electrokinetic surface potential) and the salt properties \cite{andersonprieve} but it is independent of the particle size. For non-ionic solutes $D_{p}=CD'_p$, so that $\bm v_{ph}\,\approx\,D'_p \nabla C$ where $D'_p$ is constant, see Table \ref{table}. Our consideration is independent of  the details of the dependence of $D_p$ on $C$.
In the case of electrophoresis, Smoluchowski's \cite{Smoluchowski} formula $c_E=\epsilon \zeta/(4\pi\eta_f)$ where $\epsilon$ is fluid permittivity and $\mu$ is the dynamic fluid viscosity can be used to compute the phoretic coefficient $c_{ph}$.
The behavior of $c_{ph}$ depends on the phoresis: for diffusiophoresis in ionic solutions $D_P$ can be considered constant but in the case of chemotaxis the chemical sensitivity $\chi$ can strongly depend on the local concentration of the attractant \cite{Keller,Rad}.

\begin{table*}
\begin{ruledtabular}
\begin{tabular}{|c|c|c|c|}
\renewcommand{\arraystretch}{1.6}
\bf Phoresis type        & \bf Driving gradient field $\phi$,      & \bf Phoretic velocity $\bm v_{ph}$                            &  \bf Compressibility $\nabla\cdot\bm v,\quad \quad \bm v = \bm u + \bm v_{ph}$                                                                      \\
[0.25cm] \hline \noalign{\vskip 2mm} 
Thermophoresis   & temperature $T$                 & $c_T\nabla T$                      &  $c_T\nabla^2 T+\nabla c_T\cdot\nabla T$             \\ [0.25cm] 
Diffusiophoresis & concentration of chemical    & ionic: $D_p\nabla \ln C$            & $D_p\nabla^2 \ln C$                                  \\
                 & species, salinity $C$        & nonionic: $D'_p\nabla C$             & $D'_p\nabla^2 C+\nabla D'_p\cdot\nabla C$           \\ [0.25cm] 
Electrophoresis  & electric potential $\varphi$ & $-c_E\nabla \varphi$                 & $-c_E\nabla^2 \varphi-\nabla c_E\cdot\nabla \varphi$ \\ [0.25cm] 
Chemotaxis       & chemical attractant          & $\chi(\nu)\nabla \nu$                & $\chi\nabla^2 \nu+\chi'[\nabla \nu]^2$               \\
                 & concentration, $\nu$                                              &                                                            &                                                      \\ \hline \noalign{\vskip 2mm} 
\multicolumn{4}{|c|}{Phoretic particles concentrate on a multi-fractal described by $\langle n(\bm x)n(\bm x+\bm r)\rangle=\langle n(\bm x)\rangle\langle n(\bm x+\bm r)\rangle(\eta/r)^{\Delta},\ \ \Delta>0$;}                                         \\
\multicolumn{4}{|c|}{$\Delta=\int_{-\infty}^{\infty} \langle \nabla\cdot \bm v(0)\nabla\cdot \bm v(t)\rangle dt/|\lambda_3|$ is twice the ratio of logarithmic rates of growth of infinitesimal volumes and areas.}                                                   \\
\end{tabular}
\end{ruledtabular}
\caption{\label{table} Description of clustering of phoretic particles in turbulence. The first column describes the phoretic phenomenon, the second column describes the field causing phoresis, the third column gives the phoretic velocity $\bm v_{ph}$ for motion in the gradient of corresponding field.
The fourth column provides the expression for the compressibility $\nabla \cdot \bm v$, where $\bm v = \bm u + \bm v_{ph}$.
The last row is the prediction of clustering described by the pair-correlation function of concentration $n$.
}
\end{table*}

We consider how the velocity 
of the phoretic particle changes when the carrying fluid moves macroscopically. The simplest case is that of uniform motion with time-independent velocity when the fluid moves as a whole at constant speed $\bm u$. The particle's velocity $v_p$ is then found from Galilean invariance: in the frame of the fluid the velocity is given by Eq.~(\ref{phorbs}) so that in the laboratory frame,
\begin{eqnarray}&& \bm v(t)=\bm u+c_{ph}[\bm x(t)]\nabla\phi[\bm x(t)].\label{gal}\end{eqnarray}
The particle's velocity in the fluid whose macroscopic velocity is non-constant and time-dependent can be obtained as "adiabatic version" of the above equation provided the flow changes in space and in time over scales much larger than $v_{ph}\tau_{rel}$, $d_p$ (particle diameter) and $\tau_{rel}$, respectively. At a given moment in time the fluid around the particle then has the velocity $\bm u[t, \bm x(t)]$ that changes in space only far from the particle. Due to the locality of interactions, the particle reacts as it in infinite fluid moving at constant, time-independent velocity.
If the phoretic velocity's relaxation occurs over the time during which the flow around the particle stayed constant then the resulting velocity will be given by Eq.~(\ref{gal}) with $\bm u=\bm u[t, \bm x(t)]$,
\begin{eqnarray}&&
\bm v(t)\approx \bm u[t, \bm x(t)]+c_{ph}[t, \bm x(t)]\nabla\phi[t, \bm x(t)].\label{phor}
\end{eqnarray}
The use of infinite fluid in our consideration is not a limitation, since the boundaries, breaking the Galilean invariance, are far away (the relaxation process is local so that the influence of far-away regions of the fluid is negligible).

\section{Microscopic consideration of phoresis in flows}\label{sec:microscopic}

This Section targets the derived consequence of the local Galilean invariance 
 in Eq.~(\ref{phor}) that can be obtained from microscopic considerations. These considerations provide further insight in the domain of validity of Eq.~(\ref{phor}). This will in the following be illustrated on the previously introduced example of thermophoresis. One of the microscopic approaches to this phenomenon in a fluid which is macroscopically at rest uses the Langevin equation \cite{kamp,lop}

\begin{eqnarray}&& \frac{d\bm v}{dt}=-\frac{\bm v}{\tau}+\sqrt{\frac{T[\bm x(t)] mk_B}{\tau}}\bm \xi,\label{lang}\end{eqnarray}
where $\bm \xi$ is Gaussian white noise with zero mean and pair correlation $\langle \xi_i(t)\xi_k(t')\rangle=2\delta_{ik}\delta(t-t')$. Here $m$ is the particle's mass, $k_B$ is the Boltzmann constant, $T$ is the temperature and $\tau$ is the viscous (Stokes) relaxation time. We consider spherical particles with radius $d_p$ so that $\tau=2\rho_p d_p^2/[9 \nu \rho_f]$ where $\rho_p$, $\rho_f$ are mass densities of the particle and the fluid, respectively, $\nu$ is the kinematic viscosity. The scale of spatial variations of temperature has to be much larger than $d_p$ for the description of interactions of the particle with the fluid to be describable as white noise with space-dependent amplitude (which presumes "adiabaticity" of interaction where roughly uniform temperature holds locally). When the temperature is uniform we have the usual Langevin equation describing Brownian motion of a macroscopic particle in the gas with uniform steady state distribution. In contrast, when $T$ is non-constant the steady state distribution is non-uniform because the particle accumulates in colder regions of the gas. This phenomenon can be described considering the overdamped limit $\tau\to 0$ of the Kramers equation \cite{lop} obeyed by the joint probability density $P(\bm x, \bm v, t)$ of the particle's position $\bm x$ and velocity $\bm v$,

\begin{eqnarray}&&\!\!\!\!\!\!\!\!\!\!\!\!\!\!
\partial_tP+\left(\bm v\cdot\nabla_{\bm x}\right) P=\frac{1}{\tau}\nabla_{\bm v}\cdot\left[\left(\bm v P\right)+\frac{k_B T(\bm x)}{m}\nabla_{\bm v} P\right].
\end{eqnarray}
The Maxwell distribution is the steady state solution of this equation for constant $T$ but not for space-dependent $T(\bm x)$. In the overdamped limit the spatial density $\rho(\bm x, t)=\int P(\bm x, \bm v, t)d\bm v$ obeys \cite{lop}
\begin{eqnarray}&&
\partial_t\rho=\nabla\cdot \left[\nabla \left(D(\bm x)\rho\right)\right],\ \ D(\bm x)=\frac{k_BT(\bm x)\tau}{m}.
\end{eqnarray}
Thus the probability current is $-\nabla \left(D(\bm x)\rho\right)=-\rho\nabla D-D\nabla\rho$. This has the form of the sum of a current of particles that move with average space-dependent velocity $-\nabla D(\bm x)$ and diffuse with space-dependent diffusion coefficient $D(\bm x)$. Thus temperature inhomogeneity brings particles' drift to colder regions of the fluid with velocity $-(k_B\tau/m)\nabla T$. Comparing with Eq.~(\ref{phorbs}) we can identify $\phi=T$ and $c_{ph}=-(k_B\tau/m)$.

We now consider how these formulations change when the fluid moves macroscopically with flow $\bm u(t, \bm x)$. The equation of motion Eq.\,(\ref{lang}) becomes
\begin{eqnarray}&& 
\frac{d\bm v}{dt}=-\frac{\bm v-\bm u[t, \bm x(t)]}{\tau}+\sqrt{\frac{T[\bm x(t)] mk_B}{\tau}}\bm \xi, \label{fl}
\end{eqnarray}
describing linear friction that damps differences in the particle's velocity $\bm v(t)$ and the local flow $\bm u[t, \bm x(t)]$ at the position of the particle. This equation holds provided the Reynolds number $d_p|v-u|/\nu$ based on particle's motion with respect to the flow is small and other forces such as added mass can be neglected, see the next Section and \cite{MaxeyRiley}. This equation describes thermophoresis in an external force $m\bm u[t, \bm x(t)]/\tau$. The study of the overdamped limit performed in \cite{lop} gives in this case, 
\begin{eqnarray}&&
\partial_t\rho+\bm u\cdot \nabla \rho=\nabla\cdot \left[\nabla \left(D(\bm x)\rho\right)\right].
\end{eqnarray} 
This describes the motion of particles in space with velocity,
\begin{eqnarray}&&
\bm v(t)\approx \bm u[t, \bm x(t)]-\nabla D(\bm x),\label{drft}
\end{eqnarray} which is Eq.~(\ref{phor}) with the previously derived identification $\phi=T$ and $c_{ph}=-(k_B\tau/m)$. This completes the microscopic derivation of Eq.~(\ref{phor}) that we obtained from "macroscopic" considerations based on "approximate" Galilean invariance.

In the following the condition of validity of Eq.~(\ref{drft}) will be discussed. The validity of Eq.~(\ref{fl}) demands that the smallest spatial scale $l_0$ of variations of $\bm u$ and $T$ is much larger than the particle's size. The validity of the overdamped limit demands that the time-scale of friction $\tau$ is the smallest time-scale in the problem. Thus $\tau$ has to be much smaller than the smallest time-scale of variations of flow and temperature in the particle's frame, $\bm u[t, \bm x(t)]$ and $T[t, \bm x(t)]$, respectively. These are, respectively, the smallest time-scale of turbulence (usually the Kolmogorov time, see below) and the scale $l_0/v_{ph}$ that describes the change of the fields in the particle's frame. Here $v_{ph}$ is the typical value of the phoretic velocity $c_{ph}|\nabla\phi|$ so that during time $l_0/v_{ph}$ the particle drifting through the flow will see changes in the flow around it because it enters regions with different spatial structure of the fields.

The linear relaxation of the particle's velocity to Eq.~(\ref{drft}) can be described using the effective equation 
\begin{eqnarray}&&
\frac{d\bm v}{dt}=-\frac{\bm v-\bm u[t, \bm x(t)]-c_{ph}\nabla \phi}{\tau_{rel}}, \label{mdl}
\end{eqnarray}
with $\tau_{rel}\sim\tau$, $\phi=T$ and $c_{ph}=-(k_B\tau/m)$. This effective equation captures that relaxation is linear and occurs in a time scale of order $\tau$. We propose this equation as the general model for the description of the motion of phoretic particles in flows where $c_{ph}$ and $\phi$ have to be taken in accord with the considered process. The difference between various phoretic phenomena is found in the value of $\tau_{rel}$. Clearly $\tau_{rel}$ cannot be less than the Stokes time $\tau$ however it can be much larger than $\tau$, if the time-scale of interactions $\tau_i$ (electric, chemical or others) causing the phoresis is much larger than $\tau$. In the next Section we demonstrate for spherical particles that $\tau_{rel}=\tau$ when $\tau_i\ll \tau$. Other situations have to be studied on case-by-case basis and are beyond the scope of this paper.

In the limit where $\tau$ is much smaller than the smallest time-scale of $\bm u[t, \bm x(t)]$, $\phi[t, \bm x(t)]$, Eq.~(\ref{mdl}) becomes Eq.~(\ref{drft}).
The produced conditions in terms of $l_0$ and $v_{ph}$ were considered previously. We would like to point out that Eq.~(\ref{phor}) holds beyond this model because it is based on the general principles of locality and "approximate" Galilean invariance.

\section{Fluid mechanical consideration of phoresis in flows}\label{sec:fluidmech}

This Section demonstrates how Eq.~(\ref{phor}) can be derived from fluid mechanics. We consider the motion of phoretic particle in the flow where the local neighborhood of the particle is given by approximately a constant gradient of the phoretic field. In the case considered below the field is salinity whose coupling to the flow is describable in the frame of the Boussinesq approximation. Then the assumption of approximately constant gradient states that salinity unperturbed by the particle would vary over a spatial scale much larger than the size of the particle. Similarly the flow changes over a scale much larger than the particle size. In this situation locality of interactions building up the (diffusio)phoresis implies that in the leading order the flow is a superposition of the unperturbed flow and the perturbation which is the flow that the particle would produce in the fluid at rest. That perturbation is a flow around the particle in fluid at rest when the imposed gradient of the phoretic field is the local gradient of unperturbed salinity at the position of the particle. Thus the flow perturbation produced by the particle is independent of the flow of the fluid being superposed on it (in the case where the fluid is at rest that flow perturbation is the total flow). This simple robust structure seems inevitable in the limit where the spatio-temporal variations of the unperturbed flow happen on scales much larger than the characteristic scales of the phoresis implying Eq.~(\ref{phor}). We provide the main lines of the derivation that can be turned into a detailed proof using the frame used in \cite{Maxey}.

The description of phoretic phenomena in the frame of fluid mechanics contains certain delicate points (which is the reason why the Langevin equation approach described in the previous Section has some advantages), namely it cannot be done in the frame of macroscopic no-slip boundary conditions on the surface of the particle \cite{and}. The next order corrections in Knudsen number (the parameter of fluid mechanical approximation) for the boundary condition are necessary for the fluid mechanical derivation of the phoresis \cite{and}. This causes less universality in the treatment. However for the purpose of finding how the velocity of phoretic particle changes in the presence of a macroscopic flow of the fluid the details of phoresis' derivation are less relevant.

We start from fluid-mechanical description of phoresis of small rigid particles in the fluid at rest \cite{and}. This is based on introducing finite slip velocity $\bm v^s$ on the surface of rigid particles. That violates the usual no-slip boundary condition providing effective macroscopic description of the non-trivial flow that forms near the particle's surface because of the interactions of the particle's surface with the driving gradient field $\nabla\phi$. This flow occurs in the interfacial region whose width is assumed to be much smaller than macroscopic scales and the radius $R$ of the particle (that is taken spherical for clarity). Thus the surface $S$ enclosing the particle and the interfacial region can be considered in fluid mechanical calculations as the surface of the particle. The surface flow occurs then on the particle's surface and is described by the condition that the flow outside $S$ matches that flow. It is this matching condition that is described by the slip boundary conditions. Though other ways of fluid mechanical approach to the description of phoresis were proposed recently \cite{brenner} we will stick to this more conventional one.

Below we take for definiteness the case of diffusiophoresis where $\phi$ is the salinity concentration $C(t, \bm x)$  but the calculations can be done for other phoretic phenomena similarly. The interactions occurring in the interfacial region produce on $S$ finite slip velocity of the fluid $\bm v^s$ given by solution of, 
\begin{eqnarray}&&\!\!\!\!\!\!\!\!\!\!\!\!
\bm v^s=-b\nabla C^s,
\end{eqnarray}
where $\nabla  C^s$ is the value of $\nabla C$ on the particle's surface. The coefficient $b$ is a material property
of the interface depending only on local thermodynamic conditions. It is considered as phenemenological scalar similarly as viscosity or other fluid mechanical coefficients are (it can depend on $C$ which is of no consequence below) \cite{and}. The distribution of $C$ that determines $\nabla C^s$ obeys,
\begin{eqnarray}&&\!\!\!\!\!\!\!\!\!\!\!\!
\nabla^2 C=0,\ \ \nabla C(r=\infty)=\nabla C^{\infty},\ \ {\hat n}\cdot \nabla C^s=0, \label{vs}
\end{eqnarray}
where $\nabla C^{\infty}$ is the imposed gradient not distorted by the particle, ${\hat n}$ is normal to the surface describing no flux boundary condition and the Peclet number (ratio of $R$ times phoretic velocity and the salinity diffusivity coefficient $D_S$) is considered small. We observe that $\bm v^s$ varies over the particle's surface.

Once the solution for the above problem is found providing us with $\bm v^s$ the flow of the fluid obeys the creeping flow equations with slip boundary conditions,
\begin{eqnarray}&&
-\nabla p+\nu \nabla^2 \bm u=0,\ \ \nabla\cdot \bm u=0,\label{cns0}\\&& \bm u(S)=\bm v+\bm \omega \times \bm r+\bm v^s,\label{cns}
\end{eqnarray}
where $\bm v$ is the translational and $\bm \omega$ is the angular velocity of the particle. It is assumed that the time-scale $\tau_i$ of surface interactions is much smaller than other time-scales in the problem (the Stokes time $\tau$ below) so that $\bm v^s$ can be considered instantaneously determined by $C$. We observe that though the distribution of $C$ is non-trivial distribution with typical scale $R$ its impact on $\bm u$ through buoyancy is considered to be negligible. The equations of motion read, 
\begin{eqnarray}&&
m\frac{d\bm v}{dt}=\int_S {\hat n}\cdot \sigma dS,\ \ I\frac{d\bm w}{dt}=\int_S \bm r\times (\sigma \cdot {\hat n}) dS,
\end{eqnarray}
where $\sigma$ is the fluid stress tensor, $m$ is the mass of the particle and $I=2mR^2/5$ is the moment of inertia. The particle for definiteness is considered as solid sphere with uniform density. It is found that for the special values of $\bm v=b\nabla C^{\infty}$ and $\bm \omega=0$ the total force and the torque on the particle vanish \cite{and}. These values set up after brief transients during which the particle changes its velocity under the action of finite forces from the fluid until the velocity becomes $b\nabla C^{\infty}$ and $\bm \Omega$ becomes zero. Thus the particle moves at constant velocity compensating for the surface flow so that the total force acting on it vanishes. This provides a fluid mechanical description of phoresis see \cite{and} for details. 

Transients can be described by writing the solution to Eqs.~(\ref{cns0})-(\ref{cns}) in the form of superposition of the flow with boundary conditions $b\nabla C^{\infty}+\bm v^s$ and $\bm v-b\nabla C^{\infty}+\bm \omega \times \bm r$ (similar study is performed in \cite{fo,fk} for the study of the problem of swimming in the flow - fluid mechanical problems of motion of phoretic particles and swimmers are quite similar). The former flow produces no contribution in the force or torque by construction. The other flow is that caused by a sphere that moves at the speed $\bm v-b\nabla C^{\infty}$ rotating with angular velocity $\bm \omega$. Using the corresponding Stokes force and torque we find 
\begin{eqnarray}&&
\frac{d\bm v}{dt}=-\frac{\bm v-b\nabla C^{\infty}}{\tau},\ \ \frac{d\bm w}{dt}=-\frac{10\bm \omega}{3\tau},
\end{eqnarray}
where $\tau$ is the Stokes time. Thus the relaxation to the steady phoretic drift velocity occurs at the same rate as the velocity decay in the fluid at rest.

Below we designate the flow round the particle in the fluid at rest with $\bm u^{\nabla C^{\infty}}$ and $C^{\nabla C^{\infty}}$ where the flow is considered as function of the imposed gradient of $C$.
 
We study how this consideration changes when the fluid is not at rest but rather moves with the flow that without the particle would be $\bm u_0(t, \bm x)$. The unperturbed distribution of salinity is designated by $C_0(t, \bm x)$. We consider the typical situation where the flow can be described using the Boussinesq approximation,
\begin{eqnarray}&&\!\!\!\!\!\!\!\!\!\!\!\!
\partial_t\bm u+\bm u\cdot\nabla\bm u=-\nabla p+C \bm g+\nu\nabla^2\bm u,\ \ \nabla\cdot\bm u=0, \label{basicb}\\&&\!\!\!\!\!\!\!\!\!\!\!\!
\partial_tC+\bm u\cdot\nabla C=D_S\nabla^2 C, \label{basic}
\end{eqnarray}
where $p$ is pressure divided by density. We use a rescaled field $C$ so that the buoyancy force is $C\bm g$ where $\bm g$ is the gravitational acceleration. This implies the corresponding rescaling of the diffusiophoretic coefficient below (there is no rescaling in ionic solutions where the phoretic velocity is $D_p\nabla\ln C$ though).

By definition $\bm u_0(t, \bm x)$ and $C_0(t, \bm x)$ solve Eqs.~(\ref{basicb})-(\ref{basic}) where we do not write the boundary or other driving forces if those are present (our considerations hold for quasi-stationary turbulence as well). The flow change induced by the particle is described through the boundary conditions on the particle's surface ($\bm r=\bm x-\bm x(t)$), 
\begin{eqnarray}&&\!\!\!\!\!\!\!\!\!\!\!\!
\bm u(S)=\bm v+\bm \omega \times \bm r+\bm v^s, \ \ {\hat n}\cdot\nabla C=0,\label{bc}
\end{eqnarray}
where in writing the no flux boundary condition we assume self-consistently that the difference $\bm u-\bm v$ is of order of the phoretic velocity in the fluid at rest so the convective term in the flux proportional to $(\bm u-\bm v)C$ can be neglected by smallness of the Peclet number. Since $\bm v^s$ is determined by local thermodynamic calculation \cite{and} and local thermal equilibrium holds in fluid mechanics then $\bm v^s=-b\nabla C^s$ where $C$ obeys Eqs.~(\ref{basicb})-(\ref{basic}). We use here the assumption that $\tau_i\ll \tau$ (remind that $\tau$ itself is considered much smaller than the smallest Kolmogorov time-scale of turbulence, see the previous Section). The detailed discussion of the limits of applicability of this consideration of $\bm v^s$ is beyond our scope here, see \cite{and}. 

We look for the solution of the problem set by equations (\ref{basicb})-(\ref{bc}) as the sum of the unperturbed flow $\bm u_0(t, \bm x)$  and the perturbation flow $\bm u' [t, \bm x-\bm x(t)]$ centered at the moving position of the particle $\bm x(t)$ and similarly for $C$. The perturbation flow designated by primes obeys the linearized fluid mechanical equations,
\begin{eqnarray}&&\!\!\!\!\!\!\!\!\!\!\!\!\!\!
\partial_t\bm u'\!+\![\bm u_0\!-\!\bm v]\!\cdot\!\nabla\bm u'\!+\!\bm u'\!\cdot\!\nabla \bm u_0\!=\!-\nabla p'\!+\!C' \bm g\!+\!\nu\nabla^2\bm u',\\&&\!\!\!\!\!\!\!\!\!\!\!\!\!\! \partial_tC'\!+\![\bm u_0\!-\!\bm v]\!\cdot\!\nabla C'\!+\!\bm u'\!\cdot\!\nabla C_0\!=\!D_S\nabla^2 C',\ \ \nabla\!\cdot\!\bm u'\!=\!0.\end{eqnarray}
The boundary conditions on the perturbation flow are,
\begin{eqnarray}&&\!\!\!\!\!\!\!\!\!\!\!\!
\bm u'(S)=\bm v-\bm u_0[t, \bm x(t)]+\bm v^s+\ldots,\label{bc24}
\end{eqnarray}
where the dots represents terms that are linear in $\bm x-\bm x(t)$ for $\bm x$ on the particle's surface. These terms are the first order term of the Taylor series that describes small variations of $\bm u(t, \bm x)$ over the surface of the particle (we assume that $R$ is much smaller than the smallest spatial scale of $\bm u$) and the $\bm \omega\times\bm r$ term. These terms are not relevant for the translational motion of the particle that concerns us here, cf. the study for the fluid at rest. The boundary conditions on $C'$ take the form, 
\begin{eqnarray}&&\!\!\!\!\!\!\!\!\!\!\!\!
{\hat n}\cdot\nabla C_0+{\hat n}\cdot\nabla C'=0,\ \ \nabla C'(r=\infty)=0.
\end{eqnarray}
Using smallness of Reynolds and Peclet numbers and the perturbation we find, 
\begin{eqnarray}&&
0=-\nabla_{\bm r} p'+\!\nu\nabla_{\bm r}^2\bm u',\ \  0=\!D_S\nabla_{\bm r}^2 C',\end{eqnarray}
where we dropped the buoyancy term in the equation on $\bm u'$ in consistency with the dropping of this term in the study of fluid at rest above. Further we used $\nabla C_0\sim \nabla C'$ in the vicinity of the particle for dropping $\bm u'\!\cdot\!\nabla C$ in the equation in $C'$. The solution of the equation on $C'$ is, 
\begin{eqnarray}&&
C'(\bm r, t)=C^{\nabla C_0[t, \bm x(t)]}(\bm r, t)-\bm r\cdot\nabla C_0[t, \bm x(t)],\ \ \end{eqnarray}
where we remind that $C^{\nabla C_0[t, \bm x(t)]}(\bm r, t)$ is the distribution of salinity around the phoretic particle in the fluid at rest when the imposed gradient has the value given by the gradient of the unperturbed salinity field $\nabla C_0[t, \bm x(t)]$ at the position of the particle. We find that $\bm v^s$ is as in the fluid at rest with imposed gradient $\nabla C_0[t, \bm x(t)]$. This conclusion is a consequence of the fact that in the vicinity of the particle the unperturbed profile of $C_0$ is approximately linear due to $R$ much smaller than the spatial scale of variations of $C_0$. Considering then the Stokes flow equation on $\bm u'$ with the boundary conditions (\ref{bc24}) we find the problem that we had studied already considering transients in the fluid at rest. The equation of motion is 
\begin{eqnarray}&&\!\!\!\!\!\!\!\!\!\!\!\!
\frac{d\bm v}{dt}=-\frac{\bm v-\bm u[t, \bm x(t)]-b\nabla C_0[t, \bm x(t)]}{\tau}. 
\end{eqnarray}
where where we neglect  forces other such as fluid acceleration and added mass, see \cite{Maxey}.
Using the condition that $\tau$ is much smaller than the smallest time-scale of turbulence we find that after transients on time-scale $\tau$ the motion of phoretic particle's in a flow whose spatial and temporal scales of variation are much larger than $R$ and $\tau\gg \tau_i$, respectively, is described with, 
\begin{eqnarray}&&\!\!\!\!\!\!\!\!\!\!\!\!
\bm v=\bm u_0[t, \bm x(t)]+b\nabla C_0[t, \bm x(t)],\\&&\!\!\!\!\!\!\!\!\!\!\!\! \bm u(t, \bm x)=\bm u_0(t, \bm x)+\bm u^{\nabla C_0[t, \bm x(t)]}(t, \bm x-\bm x(t)),\\&&\!\!\!\!\!\!\!\!\!\!\!\!C(t, \bm x)=C_0(t, \bm x)+C^{\nabla C_0[t, \bm x(t)]}(t, \bm x-\bm x(t))\\&&\!\!\!\!\!\!\!\!\!\!\!\!
-(\bm x-\bm x(t))\cdot\nabla C_0[t, \bm x(t)],
\end{eqnarray}
where the formulas for $\bm u(t, \bm x)$ and $C(t, \bm x)$ hold at $|\bm x-\bm x(t)|$ much smaller than the scale of variations of $\bm u_0(t, \bm x)$ and $C_0(t, \bm x)$. Though the formulas look quite cumbersome they have a simple structure described in the beginning of this Section. The flow is the sum of the unperturbed flow and the flow that would hold around the particle in the fluid at rest if the unperturbed gradient of salinity at the position of the particle was imposed. This robust structure seems inevitable when the spatial and temporal scales of the unperturbed flow are the largest spatial and temporal scales in the problem.

Finally if we use the corresponding designations for $b$ in the considered case of diffusiophoresis we have,
\begin{eqnarray}&&
\bm v(t)\approx \bm u[t, \bm x(t)]+D_p\nabla\ln C[t, \bm x(t)], \label{turb}
\end{eqnarray}
which is the special case of Eq.\,(\ref{phor}).

\section{Relevant properties of small-scale turbulence}\label{sec:smallscale}

Here we briefly discuss the properties of small-scale turbulence relevant for our study. Due to universality only quite robust properties are needed for the description: the existence of small but finite scale of smoothness, its order of magnitude, typical value of velocity gradient plus gradient's correlation time. Finer details are not required because derivations need only robust chaotic properties of the flow. This description of small-scale turbulence is incomplete both because we confine ourselves to what is needed in the study and because the properties of small-scale turbulence are still not known completely in some cases.

In general the flow field $\bm u$ and scalar field $\phi$ evolve according to

\begin{equation}
\begin{gathered}
\nabla\cdot\bm u=0 \\
\partial_t\bm u+\bm u\cdot\nabla\bm u=-\nabla p+ \nu\nabla^2\bm u +\bm f(\phi)\\
\partial_t \phi+\bm u\cdot\nabla \phi=D_\phi\nabla^2 \phi, 
\end{gathered} \label{basic4}
\end{equation}
where $D_\phi$ is the diffusivity of the field $\phi$.
Here, $\bm f(\phi)$ is a body force induced by $\phi$. If $\bm f= \bm 0$, the scalar is passive, and if $\bm f \ne \bm 0$ the scalar is active.
For the Navier-Stokes equations in the Boussinesq approximation, the body force is given by $\bm f = \hat{\rho} \bm g$ where $\hat{\rho}$ is  an equation of state linking $\phi$ to the normalized density. This could be for example $\phi$ representing the salinity $C$ [in the case of diffusiophoresis, see Eq.~(\ref{basic})] or temperature $T$ (in the case of thermophoresis).

The structure of small-scale turbulence governed by Eqs.~(\ref{basic4}) is determined by the ratio of the kinematic viscosity $\nu$ to the diffusivity $D_{\phi}$. This is referred to as the Schmidt number $\textrm Sc=\nu/D_\phi$ in case $\phi$ is a solute and as the Prandtl number $\textrm Pr$ when the scalar under consideration is the temperature. In the discussion below, $\textrm Sc$ will be used but the arguments for $\textrm Pr$ are identical.

The study below demonstrates that preferential concentration can only occur below the scale of smoothness $l_0$ of the particles' flow. The physical processes that form $l_0$ and the consequent value of that scale are not relevant in the study of clustering. This is because clustering holds in arbitrary smooth flow with finite (Lagrangian) correlation time of the gradients. This guarantees that the motion of particles below $l_0$ can be described as motion in the smooth flow with linear spatial profile determined by the matrix of velocity gradients $\nabla_i v_k$. The finite correlation time of that matrix is used for predicting that the motion of small volumes of particles at large times is determined by lots of independent random deformations by $\nabla \bm v$ at different times, cf. the discussion of Eq.~(\ref{defintion}). Thus for the purposes of deriving the clustering at small scales the only relevant property of Eqs.~(\ref{basic4}) is smoothness below the small but finite scale $l_0$.

In order to perform a comparison with the experiment we do need the scale $l_0$. The smallest scale of spatial variations of $\bm u$ is the Kolmogorov length scale $\eta=\sqrt{\nu/\lambda}$. Here $\lambda=\sqrt{\epsilon/\nu}$ is the typical value of velocity gradients of turbulence (inverse Kolmogorov time-scale) and $\epsilon$ is the turbulent kinetic energy dissipation rate per unit volume \cite{Frisch}. This is the scale at which the non-linear advective acceleration $\bm u\cdot\nabla\bm u$ and the viscous terms $\nu \nabla^2 \bm u$ in Eq.~(\ref{basic4}) balance each other when $\bm f(\phi)$ is negligible (viscous scale of the Navier-Stokes turbulence). In a stably stratified turbulent flow, such as the gravity current studied in the experimental section (Section\,\ref{section4}), the buoyancy force can be neglected below the so-called Ozmidov scale \cite{odier}. In our experiment this scale is greater than the Taylor microscale $\lambda_T$ (Table\,\ref{tab:flowprop}), so at the viscous scale, the scalar field $\phi$ is passive.

The counterpart of $\eta$ for the scalar field $\phi$, i.e.\  the scale $l_d$ at which $\bm u\cdot\nabla \phi$ and $D_\phi\nabla^2 \phi$ in Eq.~(\ref{basic4}) balance, depends on $\textrm Sc$. In the case of $\textrm Sc\gtrsim 1$ this is the Batchelor scale, $l_d=\sqrt{D_{\phi}/\lambda}$. When $\textrm Sc\gg 1$ the Batchelor scale is much smaller than $\eta$. The flow in the range $l_d\ll r\ll \eta$ is differentiable with fluctuations of velocity at scales $r$ of order $\lambda r$. The variance of $\phi$ is cascaded by smooth flow from $\eta$ to $l_d$ where it is stopped by diffusion \cite{Batchelor}.

The considered case of large $\textrm Sc$ is of practical relevance in typical oceanic applications and the experiment described in Section \ref{section4} where $\textrm Sc=\nu/D_S\sim 10^3$. Here we substitute $D_{\phi}$ for the general case (Eqs.\,\ref{basic4}) by the salt diffusion coefficient $D_S$ which is what determines the size of the Batchelor scale in oceanic flows. The Batchelor scale $l_d=\sqrt{D_S/\lambda}$ is the scale where diffusion balances the local shrinking of filaments of salinity by gradients of the flow and this is the typical scale of variations of $\phi$ in oceanic flows. In this case the correlation scale $l_d$ of gradients of $\phi$ is much smaller than that of gradients of $\bm u$. The correlation scale of gradients of the flow of particles in Eq.~(\ref{phor}) is determined by the Batchelor scale $l_d$ and not the Kolmogorov scale $\eta$ so that $l_0=l_d$.

In the case $\textrm Sc\ll 1$ the scale at which $\bm u\cdot\nabla \phi$ and $D_\phi\nabla^2 \phi$ in Eqs.~(\ref{basic4}) balance is $D_{\phi}^{3/4}\epsilon^{-1/4}$ where we use Kolmogorov-Obukhov scaling in the inertial range. This scale is larger than the Kolmogorov scale that can be written in the form $\eta=\nu^{3/4}\epsilon^{-1/4}$. Thus in this case $l_0=\eta$.

We designate that below the smoothness scale of $\bm v$, $l_0$ can be generally written as $l_0=\min[\eta, l_d]$. \\

\section{Phoretic clustering in turbulence}
\label{section:2}

In this Section we demonstrate theoretically that particles drifting due to phoresis cluster in turbulence. We introduce a universal framework for different phoretic phenomena including thermophoresis, electrophoresis, chemotaxis and diffusiophoresis, see Table\,\ref{table}. Other cases where our predictions have potential applications are barophoresis and pycnophoresis, see \cite{baro} and references therein. This universality is possible because clustering of phoretic particles in turbulence is a direct consequence of the fractality of the distribution of particles in weakly compressible random flows and local Galilean invariance of fluids.

As argued in the previous Section, the motion of a phoretic particle with coordinate $\bm x(t)$ in a turbulent flow $\bm v(t, \bm x)$ is governed by
\begin{eqnarray}&&
\frac{d\bm x}{dt}=\bm v[t, \bm x(t)],\ \ \bm v=\bm u+c_{ph}\nabla\phi,\label{descr}
\end{eqnarray}
see Eq.~(\ref{phor}). The condition of validity of this description is that the scale of spatial variations of the field $\phi$ is much larger than the particle size. Hence, $\lambda\tau_{rel}\ll 1$ and $v_{ph}\tau_{rel}\ll l_0$ where $v_{ph}$ is the typical value of the phoretic velocity $c_{ph}|\nabla\phi|$.

We consider the case where the particle flow has weak compressibility so the flow divergence is much smaller than the typical value of the gradients of turbulence,
\begin{eqnarray}
&& |\nabla\cdot (c_{ph}\nabla\phi)|\ll \lambda. \label{condition1}
\end{eqnarray}
Using that $|\nabla\cdot (c_{ph}\nabla\phi)|\sim |c_{ph}\nabla\phi|/l_0\sim v_{ph}/l_0$ we find that the condition of weak compressibility is,
\begin{eqnarray}
&& v_{ph} \ll \lambda l_0. \label{condition11}
\end{eqnarray}
We observe that  ${|\nabla\cdot (c_{ph}\nabla\phi)|}/{\lambda}\approx(v_{ph}\tau_{rel}/l_0) (\lambda\tau_{rel})^{-1}$;
that is, the validity of conditions (\ref{condition1})-(\ref{condition11}) is determined by which of the two small numbers $\lambda\tau_{rel}$, $v_{ph}\tau_{rel}/l_0$ is smaller. This depends specifically on the considered case - namely the constants and the gradients of the phoretic field $\phi$. For thermophoresis in the case of non-small $\textrm Pr$ we have $\lambda \tau_{rel}\sim \lambda\tau$, $v_{ph}\tau_{rel}/l_0\sim (k_B \tau/m)\tau\nabla T/\eta$. The ratio $(v_{ph}\tau_{rel}/l_0) (\lambda\tau_{rel})^{-1}\sim (k_B \tau/m)\nabla T/\lambda\eta $ can be small or large. Considering constants fixed depending on the strength of the gradients of temperature we can have situations of small or non-small compressibility. In the case of diffusiophoretic particles in oceans whose typical parameters are provided below the assumption of weak compressibility holds well.

The weak compressibility condition shows that during the correlation time $\lambda^{-1}$ of small-scale eddies the particle deviates from the trajectories of the fluid particles by a distance much smaller than $l_0$ (the deviation of trajectories is caused by the drift velocity $v_{ph}$). Thus the gradients of the flow in the frame of the particle change over the same Kolmogorov time-scale $\lambda^{-1}$ as the gradients in the frame of the fluid particle. We will use the fact that the correlation time of $\nabla \bm v$ is the Kolmogorov time-scale in the following.

The weak compressibility of the particle flow implies that the particle distribution in space can be described completely using the universal description of particle distribution statistics in weakly compressible flow, introduced in \cite{fouxon1,FFS}. It was demonstrated in \cite{fouxon1} that in the steady state the particles concentrate on a random time-dependent multi-fractal in space. The statistics of the particle concentration field $n(t, \bm x)$ is log-normal so that the correlation functions derive from the pair correlation function
\begin{eqnarray}&&
\langle n(0)n(\bm r)\rangle=\langle n\rangle^2\left(\frac{\eta}{r}\right)^{\Delta},\ \ r\ll l_0,\label{par}\\&&
\langle n(\bm x_1)n(\bm x_2)..n(\bm x_k)\rangle=\prod_{i>k}\langle n(\bm x_i)n(\bm x_k)\rangle, \label{lognorm}
\end{eqnarray}
where $\Delta$ is the correlation co-dimension of the fractal that is given by
\begin{eqnarray}&& \!\!\!\!\!\!\!\!\!
\Delta=\frac{1}{|\lambda_3|}\int_{-\infty}^{\infty} \langle \nabla\cdot[c_{ph}\nabla \phi](0)\nabla\cdot[c_{ph}\nabla \phi](t)\rangle dt\!.\label{Delta}
\end{eqnarray}
Here $|\lambda_3|$ is the Lyapunov exponent associated with the growth exponent of infinitesimal areas, see below.
The averages in Eq.~(\ref{par})-(\ref{Delta}) are spatial,
\begin{equation}
\langle n(0)n(\bm r)\rangle=\int n(\bm x)n(\bm x+\bm r) \frac{d\bm x}{\Omega},
\end{equation}
\begin{multline}\label{func}
\langle \nabla\cdot[c_{ph}\nabla \phi](0)\nabla\cdot[c_{ph}\nabla \phi](t)\rangle= \\
\int\nabla\cdot[c_{ph}\nabla \phi](0, \bm x) 
\nabla\cdot[c_{ph}\nabla \phi][t, \bm q(t, \bm x)]\frac{d\bm x}{\Omega},
\end{multline}
where $\Omega$ is the total volume which is set below to one. We introduced the spatial Lagrangian trajectories of the fluid particles labeled by their position at $t=0$
\begin{eqnarray}&&\!\!\!\!\!\!\!\!\!
\partial_{t}\bm q(t, \bm x)=\bm u[t, \bm q(t, \bm x)],\ \ \bm q(t=0, \bm x_0)=\bm x_0.\label{lagr0}
\end{eqnarray}
The described predictions hold for spatially uniform statistics of turbulence provided $\Delta\ll 1$. The case of inhomogeneous statistics is considered in the next Section.

Equation~\eqref{Delta} is the main result of this Section: phoretic particles form a multifractal in turbulent flow with log-normal statistics determined by Eqs.~(\ref{par})-(\ref{Delta}), see Table \ref{table}. Logically, this is what we will base further calculations and our experimental validation in Section \ref{section4} on.

In the following we clarify and discuss these results. The correlation co-dimension $\Delta$ coincides with twice the Kaplan-Yorke co-dimension $D_{KY}$
\begin{eqnarray}&&
\Delta=2D_{KY},
\end{eqnarray}
whose definition \cite{KY} in the case of weak compressibility reduces to the ratio of logarithmic growth rates of infinitesimal volumes $\delta V$ and areas $\delta A$ of particles \cite{fphl}. The simplest  definition of $\delta A$ is found considering the area of a triangle formed by three particles. Similarly, $\delta V$ is defined by four particles in close proximity that form a tetrahedron. We have
\begin{eqnarray}&& \!\!\!\!\!\!\!\!\!\!\!\!\!\!\!
D_{KY}\!=\!\left|\lim_{t\to\infty}\!\frac{1}{t}\ln\left(\frac{\delta V(t)}{\delta V(0)}\right)\right|\left[\lim_{t\to\infty}\!\frac{1}{t}\ln\left(\frac{\delta A(t)}{\delta A(0)}\right)\right]^{-1}\!\!.\label{codimension}
\end{eqnarray}
The weak compressibility causes $D_{KY}$ to be much smaller than unity: for incompressible flow the volumes are conserved but the areas grow with finite exponent when the flow is chaotic (which the turbulent flow below the Kolmogorov scale is). The limits in $D_{KY}$ hold deterministically involving no averaging because the limiting rates coincide for different initial positions of volumes and areas \cite{reviewt,FB}. The limit for the volume is called the sum of the Lyapunov exponents $\lambda_i$,
\begin{eqnarray}&&
\sum \lambda_i=\lim_{t\to\infty}\frac{1}{t}\ln\left(\frac{\delta V(t)}{\delta V(0)}\right)=\nonumber\\&&-\int_0^{\infty}\langle \nabla\cdot[c_{ph}
\nabla \phi](0) \nabla\cdot[c_{ph}\nabla \phi](t)\rangle dt,
\end{eqnarray}
where we use the formula for $\sum\lambda_i$ derived in \cite{FF} and $i=1..3$ being the three different spatial directions.
The three Lyapunov exponents are ordered such that $\lambda_1 > \lambda_2 > \lambda_3$, and we note that for fluid particles ($\nabla \cdot \bm u = 0$) we would have $\sum \lambda_i = 0$.
 The negative sign of $\sum \lambda_i$ indicates that particles migrate to regions with negative flow divergence, see the discussion in the next Section. If we consider four infinitesimally separated particles, the volume $\delta V(t)$ of the tetrahedron that they form will decrease at large times exponentially at the rate $\sum\lambda_i$ identical for different initial positions of the particles and different initial times. Since the correlation time of $\nabla\cdot \bm v$ is the Kolmogorov time-scale then we find,
\begin{eqnarray}&& \!\!\!\!\!\!\!\!\!
\frac{\sum \lambda_i}{\lambda}\sim \frac{v_{ph}^2}{l_0^2\lambda^2}\ll 1, \label{magn}
\end{eqnarray}
where we used Eq.~(\ref{condition11}). The logarithmic rate of growth of infinitesimal areas $\delta A$ is non-zero for fluid particles so that considering the smallness of the phoretic component of the flow we can use the Lagrangian trajectories of the fluid particles in Eq.~(\ref{codimension}) (this approximation would fail for $\sum \lambda_i$ because that is zero for turbulence). For fluid particles volumes are conserved such that the growth exponent of infinitesimal areas coincides with the third Lyapunov exponent (see the next Section),
\begin{eqnarray}&&\!\!\!\!\!\!\!\!
|\lambda_3|=\lim_{t\to\infty}\frac{1}{t}\ln\left(\frac{\delta A(t)}{\delta A(0)}\right), \label{lambda_3}
\end{eqnarray}
where the simplest configuration that determines $|\lambda_3|$ is the triangle formed by three infinitesimally close fluid particles. The area of the triangle growth becomes at large times deterministic with an exponent given by $|\lambda_3|$ identical for all triangles. There is no simple way of writing $|\lambda_3|$ in terms of correlation functions of turbulence so it has to be considered as phenomenological positive quantity of order $\lambda$ so that,
\begin{eqnarray}&& \!\!\!\!\!\!\!\!\!\!\!\!\!\!\!
D_{KY}\!=\frac{|\sum \lambda_i|}{|\lambda_3|}\sim \frac{\sum \lambda_i}{\lambda}\sim \frac{v_{ph}^2}{l_0^2\lambda^2}\ll 1. \label{dky_lnvt}
\end{eqnarray}
where we used Eq.~(\ref{magn}). This formula provides a simple way of estimating the correlation dimension in practice. We stress that the weakness of compressibility implies smallness of the fractal co-dimension $D_{KY}$ but not of fluctuations of concentration that can be arbitrarily large.

In the following we comment on the validity of Eq.~(\ref{Delta}). The original formula for the average in the pair-correlation function (Eq.\,\ref{func}) in $\Delta$ does not involve the trajectories of the fluid particles $\bm q(t, \bm x)$ but the trajectories of the phoretic particles $\bm x(t, \bm x)$,
\begin{eqnarray}&&\!\!\!\!\!\!\!\!\!
\partial_{t}\bm x(t, \bm x)=\bm v[t, \bm x(t, \bm x)],\ \ \bm x(t=0, \bm x_0)=\bm x_0. \label{phortr}
\end{eqnarray}
defined by $\bm v$ and not $\bm u$, see \cite{fouxon1}. However condition (\ref{condition11}) implies that ($l_0\leq \eta$)
\begin{eqnarray}
&& \frac{v_{ph}}{\lambda\eta}\leq \frac{v_{ph}}{\lambda l_0} \ll 1,
\end{eqnarray}
that is, the typical value of the phoretic velocity is much smaller than the typical value $\lambda \eta$ of the turbulent velocity at the scale $\eta$. Thus during the Lagrangian correlation time $\lambda^{-1}$ of turbulent velocity gradients in the fluid particle's frame the phoretic particle deviates from the fluid particle by a distance much smaller than the correlation scale of the gradients $\eta$ that is $|\bm q(t=\lambda^{-1}, \bm x)-\bm x(t=\lambda^{-1}, \bm x)|\ll\eta$. Thus over the correlation time $\lambda^{-1}$ which determines the time integral in Eq.~(\ref{Delta}) the gradients in the fluid's and phoretic particle's frames coincide so we can use $\bm q(t, \bm x)$ instead of $\bm x(t, \bm x)$ in Eq.~(\ref{func}).


\section{Preferential concentration in inhomogeneous turbulence}\label{derivation}
In this Section we derive the pair-correlation function $\langle n(\bm x)n(\bm x+\bm r)\rangle$ of the concentration field $n(t, \bm x)$ of particles in the case where the statistics of turbulence is inhomogeneous. We find a universal formula for pair correlations of particles in inhomogeneous weakly compressible random flow. Though different inhomogeneities of the flow produce different spatial profiles $n_0(\bm x)$ of average concentration $\langle n(\bm x)\rangle=n_0(\bm x)$ we demonstrate that fluctuations of normalized concentration,
\begin{eqnarray}&& \!\!\!\!\!\!\!\!\!
{\tilde n}(t, \bm x)=\frac{n(t, \bm x)}{n_0(\bm x)},
\end{eqnarray}
obey universal statistics. These coincide with those of concentration for spatially uniform statistics described in the previous Section. We find that ${\tilde n}(t, \bm x)$ has log-normal statistics (\ref{lognorm}) which are completely determined by the pair-correlation function
\begin{eqnarray}&&
\langle {\tilde n}(\bm x){\tilde n}(\bm x+\bm r)\rangle=\left(\frac{l_0}{r}\right)^{\Delta(\bm x+ \frac{\bm r}{2})} \ r\ll l_0,\label{parcor}
\end{eqnarray}
where the difference from the spatially uniform case is that $\Delta$ is a function of the coordinate that reflects inhomogeneity of the velocity statistics,
\begin{multline}\label{inh}
\Delta(\bm x)\,=\\
\,\frac{1}{|\lambda_3(\bm x)|}\!\int_{-\infty}^{\infty}\!\!\!\! dt \left\langle \nabla\cdot[c_{ph}\nabla \phi](0)\nabla\cdot[c_{ph}\nabla \phi](t)\right\rangle,
\end{multline}
where the inhomogeneous time correlation function in the integrand is determined using trajectories that issue from $\bm x$.
In this way, describing the statistics of concentration of phoretic particles in inhomogeneous turbulence reduces to the problem of determining the concentration profile $n_0(\bm x)$ and $\Delta(\bm x)$ provided the weak compressibility condition (\ref{condition11}) holds. In this work we concentrate on deriving Eq.\,(\ref{parcor}) considering $n_0(\bm x)$ and $\Delta(\bm x)$ as phenomenological fields determined by the details of statistics of turbulence. The study of how $n_0(\bm x)$ can be obtained from the statistics of turbulence is undertaken in \cite{Peter}.

The pair-correlation function of concentration describes the probability to find a particle at distance $\bm r$ from a particle at $\bm x$ so that it enters the collision kernel determining the rate of coagulation of colloids having direct practical applications. In inhomogeneous cases the probability depends both on $\bm r$ and $\bm x$. Thus the statistics are defined by time averaging,

\begin{multline}
\langle n(\bm x)n(\bm x+\bm r)\rangle=\\
\lim_{t_0\to\infty}\frac{1}{t_0}\int_0^{t_0} n(t, \bm x) n(t, \bm x+\bm r)dt,
\end{multline}
\begin{equation}
n_0(\bm x)=\langle n(\bm x)\rangle=\lim_{t_0\to\infty}\frac{1}{t_0}\int_0^{t_0} n(t, \bm x)dt .
\end{equation}

The pair-correlation function can be obtained by multiplying the probability $\langle n(\bm x)\rangle$ of finding a particle at $\bm x$ by the conditional probability $P(\bm x|\bm r)$ of finding a particle at $\bm x+\bm r$ given that there is a particle at $\bm x$ (here the angular brackets stand for temporal averaging at fixed spatial positions, see definitions below). When $\bm r$ becomes large the location of the particle at $\bm x$ does not influence the probability $P(\bm x|\bm r)$ of finding a particle at $\bm x+\bm r$ so that $P(\bm x|\bm r)\approx \langle n(\bm x+\bm r)\rangle$ and $\langle n(\bm x)n(\bm x+\bm r)\rangle\approx \langle n(\bm x)\rangle\langle n(\bm x+\bm r)\rangle$. Thus at large separations the pair-correlation function decomposes to the product of averages describing independence of concentration fluctuations at separated points. In contrast, when $r\to 0$ there is an increase in $P(\bm x|\bm r)$ reflecting particles clustering together in preferred regions of the flow - preferential concentration. It is this amplification factor,

\begin{multline}\label{factor}
f(\bm x, \bm r)\,=\,\frac{\langle n(\bm x)n(\bm x+\bm r)\rangle}{\langle n(\bm x)\rangle \langle n(\bm x+\bm r)\rangle}=\langle {\tilde n}(\bm x){\tilde n}(\bm x+\bm r)\rangle, 
\end{multline} for which we derive a closed-form expression in this Section. This factor is a "proper correlation": if $n_0(\bm x)$ is larger in certain regions of space then particles will tend to go to that region independently of the behavior of other particles so the product $n(\bm x)n(\bm x+\bm r)$ will be larger there trivially. Our derivation holds for arbitrary weakly compressible flow so that it can be used for all the phoretic phenomena described in the previous Section, inertial particles in turbulence at small Stokes or Froude numbers \cite{fouxon1,fphl} or other cases.

\begin{figure}
\includegraphics[width=8.0 cm,clip=]{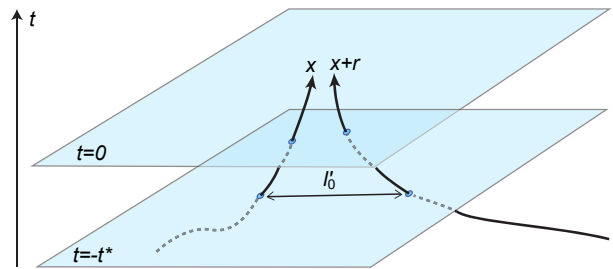}
\caption{Illustration of positive correlations of phoretic particles in turbulence. 
When random turbulent transport brings two particles to a distance $l_0$ at time $-t^*$, their common motion in the predominantly negative divergence of the flow creates effective attraction between the particles.
The pair-correlation $\langle n(\bm x)n(\bm x+\bm r)\rangle$ is the probability $\langle n(\bm x)\rangle \langle n(\bm x+\bm r)\rangle$ to randomly get close by distance $l_0$ times the increase factor (see Eq.\,\ref{constr}) due to common motion in the predominantly negative divergence.}
\label{fig1}\end{figure}

The reasons why turbulence increases the probability of two particles to get close can be understood from the fact that on average the divergence of velocity in the particle's frame is negative $\langle \nabla\cdot\bm v[t, \bm x(t, \bm x)]\rangle<0$. Particles tend to go to regions where the divergence is negative so in the particle's frame the divergence is mostly negative. Thus when two particles transported by turbulence are randomly brought below the "minimal correlation length" of velocity divergence $l'_0$ they start moving in the same divergence which is typically negative. Motion in common divergence causes the particles to preferentially approach each other producing $f(\bm x, \bm r)>1$, see Fig. {\ref{fig1}}. Here $l'_0$ is the largest scale over which $\nabla\bm v$ can be considered constant which can be taken one order of magnitude smaller than $l_0$. We will demonstrate that there is no correlation of concentration fluctuations at $l'_0$.

We consider an increase in the probability of two particles carried along by turbulence to approach each other at distance $\bm r\ll l'_0$ at the time of observation $t=0$ which is described by the pair-correlation function. We can separate the history of the particles' motion in space at $t<0$ to times $t<t^*$ when the particles' separation $\bm r(t)$ was larger than $l'_0$ and times $t^*<t<0$ where $\bm r(t)<l'_0$, see Fig. {\ref{fig1}}. The particles moved in uncorrelated divergences of the flow at $t<t^*$ so there was no preference to getting closer or further (the residual power-law correlations in the inertial range have small but finite value which we study below; these are not relevant for finding the leading order term here). The increase in probability is built in the last period of motion in the common velocity divergence. This can be described by using the continuity equation
\begin{eqnarray}&&
\partial_t n+\nabla\cdot(n\bm v)=0, \label{continuity}
\end{eqnarray}
which has the solution ($n(\bm x)=n[t=0, \bm x]$)
\begin{multline}
n(\bm x)=\\
n\left[t, \bm x\left(t, \bm x\right)\right]\exp\left[-\int_t^0 \nabla\cdot\bm v[t', \bm x(t', \bm x)]dt'\right],\label{cont}
\end{multline}
where $\bm x(t, \bm x)$ is the particle trajectory that passes at $t=0$ through the point $\bm x$, see Eq.~(\ref{phortr}).
Taking the average of the product of $n(\bm x)$ and $n(\bm x+\bm r)$, we find that
\begin{eqnarray}&&\!\!\!\!\!\!\!\!
\langle n(\bm x)n(\bm x+\bm r)\rangle\,=\,\biggl\langle n[t, \bm x(t, \bm x)]n[t, \bm x(t, \bm x+\bm r)]\label{pairc}\\&&\!\!\!\!\!\!\!\!\exp\left[-\!\!\int_t^0 \!\!\!dt'\left(\nabla\cdot\bm v[t', \bm x(t', \bm x)]\!+\!\nabla\cdot\bm v[t', \bm x(t', \bm x\!+\!\bm r)]\right)\right]\biggr\rangle.\nonumber
\end{eqnarray}\\
We demonstrated that pair correlations form when the distance between the particles is much less than $l'_0$. We consider $r\ll l'_0$ and track the trajectories $\bm x(t, \bm x)$ and $\bm x(t, \bm x+\bm r)$ back in time in order to determine the positive correlation accumulated during the times when the distance $\bm r(t)=\bm x(t, \bm x+\bm r)-\bm x(t, \bm x)$ between the trajectories was less than $l'_0$.

We briefly sketch the properties of evolution of distances below the Kolmogorov scale in the following, see \cite{Oseledec,Goldhirsch,FB,reviewt} for details.
The separation velocity is linear in $\bm r$ at $r<l'_0$ because the particles' velocity difference can be approximated by separation $\bm r$, times the local flow gradient. Thus the separation below $l'_0$ behaves exponentially and is characterized by a positive exponent describing chaoticity of motion of particles below $l'_0$,
\begin{eqnarray}&&\!\!\!\!\!\!\!\!
\lim_{t\to-\infty} \frac{1}{|t|}\ln\left(\frac{r(t)}{ r(0)}\right)\approx |\lambda_3|, \label{sep}
\end{eqnarray}
where $\lambda_3$ is the third Lyapunov exponent of the fluid particles in turbulence. Thus at large times the growth of distances between trajectories back in time is a deterministic exponential growth with exponent $|\lambda_3|$. This exponent can be seen in the forward in time evolution of an infinitesimal ball of fluid particles with size $r_0$ much less than $\eta$. Turbulence deforms the ball to an ellipsoid whose axes behave exponentially \cite{Oseledec,Goldhirsch,FB,reviewt}. The major axis increases as $r_0\exp[\lambda_1 t]$ where $\lambda_1>0$ is the principal Lyapunov exponent. The minor axis decreases as $r_0\exp[\lambda_3 t]$ where $\lambda_3<0$ is the third Lyapunov exponent. The exponential evolution of the intermediate axis $r_0\exp[\lambda_2 t]$ is determined by the volume conservation condition $\lambda_1+\lambda_2+\lambda_3=0$. Then the growth of the distance between two fluid particles is given by $\lambda_1$,
\begin{eqnarray}&&\!\!\!\!\!\!\!\!
\lim_{t\to\infty} \frac{1}{t}\ln\left(\frac{ r(t)}{ r(0)}\right)\approx \lambda_1,
\end{eqnarray}
which is the forward in time counterpart of Eq.~(\ref{sep}). This holds at times much larger than the correlation time $\lambda^{-1}$ of flow gradients which determine the velocity difference of close particles constituting a form of ergodic theorem or the law of large numbers \cite{Oseledec,Goldhirsch,FB,reviewt}. When evolution is time-reversed the major axis of the ellipsoid starts to grow at the exponent $|\lambda_3|$. Thus it is $|\lambda_3|$ that gives the logarithmic rate of separation of fluid particles back in time, see Eq.~(\ref{sep}). The rate of separation of phoretic particles approximately coincides with $|\lambda_3|$ because the phoretic component of velocity is small. Thus Eq.~(\ref{sep}) holds for both fluid and phoretic particles.

We conclude that when $r\to 0$ the time
\begin{eqnarray}&&\!\!\!\!\!\!\!\!
t^*=\frac{1}{|\lambda_3|}\ln\left(\frac{l'_0}{r}\right),\label{tst}
\end{eqnarray}
that exponentially diverging trajectories $\bm x(t, \bm x)$ and $\bm x(t, \bm x+\bm r)$ spend below $l'_0$ grows logarithmically getting infinite at $r=0$. This is because turbulence is smooth below $\eta$. The pair-correlations form at $r=0$ for infinite time causing the divergence of $\langle n^2(\bm x)\rangle$, see below.

Inclusion of small but finite Brownian motion of the particles would cause the trajectories with $r=0$ to diverge in finite time. The separation occurs diffusively (with linearly growing in time dispersion) until the diffusive scale $\sqrt{\kappa/|\lambda_3|}$ is reached. Starting from this scale trajectories separate because of difference of local velocities and diffusion can be neglected \cite{FB,FFS,reviewt}. Correspondingly the correlation function is cut off by diffusion at the diffusive scale below which the correlation is roughly constant, see \cite{FFS,reviewt}. We assume throughout the paper that this diffusive scale $\sqrt{\kappa/|\lambda_3|}$ is much smaller than $l_0$ so that there is a range of separations where the considered purely fluid mechanical trajectories hold. We consider scales higher than $\sqrt{\kappa/|\lambda_3|}$ and neglect diffusion.

If turbulence is inhomogeneous then it is necessary to refine the considerations because the rate of separation $\lambda_3$ in this case depends on the position of the particles at $t=0$. We consider the case which is typical in practice where the center of mass of the separating pair of particles stays in the region where the turbulent statistics is approximately uniform during the time interval $-t^*<t<0$. In other words the scale $L$ of inhomogeneity of turbulent statistics is assumed to be much larger than the typical distance $|\bm x(-t^*, \bm x)-\bm x|$. Since $t^*$ diverges when $l\to 0$, see Eq.~(\ref{tst}), then this implies that we consider not too small $l$. Then we can define
\begin{eqnarray}&&\!\!\!\!\!\!\!\!
\frac{1}{t^*}\ln\left(\frac{|\bm x(-t^*, \bm x+\bm r)-\bm x(-t^*, \bm x)|}{r}\right)\approx |\lambda_3(\bm x)|, \label{defintion}
\end{eqnarray}
that holds provided $\ln(l'_0/r)\gg 1$ or $|\lambda| t^*\gg 1$, see Eq.~(\ref{tst}). The inequality guarantees that the LHS is the sum of $N\sim |\lambda| t^*\gg 1$ independent random variables divided by $N$ so the law of large numbers holds defining a unique realization-independent function $\lambda_3(\bm x)$. In practice the logarithm is never too large so our consideration is an asymptotic study which then is continued to the physical range of parameters - the formulas derived under condition $\ln(l'_0/r)\gg 1$ hold when $l'_0/r\gg 1$ as can be proved applying the cumulant expansion theorem to Eq.~(\ref{pairc}) in the steady state limit $t\to -\infty$. Here we sketch the proof, see details in \cite{fouxon1}. We set with no loss the initial condition at time $t$ to a constant, $n(t, \bm x)=\langle n\rangle$. Then the cumulant expansion theorem gives
\begin{eqnarray}&&\!\!\!\!\!\!\!\!
\ln \langle n(\bm x)n(\bm x+\bm r)\rangle\,=\sum_{k=1}^{\infty}\frac{1}{k!}\lim_{t\to-\infty} \nonumber\\&&\!\!\!\!\!\!\!\!
\biggl\langle\left[-\!\!\int_t^0 \!\!\!dt'\left(\nabla\cdot\bm v[t', \bm x(t', \bm x)]\!+\!\nabla\cdot\bm v[t', \bm x(t', \bm x\!+\!\bm r)]\right)\right]^k\biggr\rangle_c,\nonumber
\end{eqnarray}
where $c$ stands for cumulant. Opening the brackets one finds correlation functions that have finite steady state limit $t\to-\infty$.  have the  at scales $r$ much smaller than the correlation scale $l'_0$ of the gradients. Using corresponding asymptotic form of these functions at $l'_0/r\gg 1$ one recovers the result obtained previously under the more stringent condition $\ln(l'_0/r)\gg 1$. 

When we consider a decrease of $r$ the time $t^*$ increases indefinitely (it grows infinite logarithmically in $r$). When $t^*$ gets large the displacement $|\bm x(-t^*, \bm x)-\bm x|$ will reach $L$ causing fluctuations in the LHS of Eq.~(\ref{defintion}) caused by trajectories' explorations of spatial regions with different statistics of turbulence. Further increase in $t^*$ will produce the trajectory that explores the whole volume of the flow so that the LHS will become constant independent of the coordinate. This constant is the rigorous mathematical definition of the Lyapunov exponent that is however of little practical use when a large volume is studied.

Thus the fluctuations of concentration at not too small $r$ are determined by $\lambda_3(\bm x)$ characterizing the local statistics of turbulence. When smaller $r$ are studied the inhomogeneity of the turbulent statistics would cause changes in the LHS of Eq.~(\ref{defintion}) as the center of mass of the particles explores regions of the flow larger than $L$ over which the statistics is inhomogeneous. These scales are not relevant in the common situation when $L$ is much larger than $l_0$ and will not be studied in this work.

We observe that we consider the time $t^*$ to separate from initial (or rather final) distance $r$ to $l'_0$ as deterministic quantity. This neglects the fluctuations of finite-time Lyapunov exponents (large deviations \cite{FB,review}). Consistent inclusion of the fluctuations demonstrates that those can be neglected because weakness of compressibility causes the averages to be determined by the most probable $\lambda_3$ and not the large deviations \cite{fouxon1}.

We consider Eq.~(\ref{pairc}) at $t=-t^*$. The average in the RHS contains both averaging over times smaller than $-t^*$ and the times of formation of pair correlations $-t^*<t<0$. For separating these contributions we observe that the condition of weak compressibility (\ref{condition1}) implies that time integral of $\nabla\cdot\bm v[t, \bm x(t, \bm x)]$ over times of order of the correlation time $\lambda^{-1}$ of $\nabla \cdot \bm v$ is much less than one (thus over these time-scales the concentration is conserved in the particle's frame,
\begin{eqnarray}&&\!\!\!\!\!\!\!\!\!\!\!\!\!
n(\bm x)\approx n\left[t, \bm x\left(t, \bm x\right)\right],\ \  \lambda |t|\lesssim 1,
\end{eqnarray}
which is another way of describing weak compressibility of the flow). Neglecting the contribution of times in $\lambda^{-1}-$vicinity of $-t^*$, we find that the concentration factors in the first line of Eq.~(\ref{pairc}) are independent of the exponential in the last line dependent on the "future flow":

\begin{eqnarray}&&\!\!\!\!\!\!\!\!
\langle n(\bm x)n(\bm x+\bm r)\rangle\!\approx \!\left\langle n[-t^*, \bm x(-t^*, \bm x)]n[-t^*, \bm x(-t^*, \bm x\!+\!\bm r)]\right\rangle\nonumber\\&&\!\!\!\!\!\!\!\!\left\langle\exp\left[-2\int_{-t^*}^0 dt\nabla\cdot\bm v[t, \bm x(t, \bm x+\frac{\bm r}{2})]\right]\right\rangle,\nonumber
\end{eqnarray}\\
where we used that $(\nabla\cdot\bm v[t, \bm x(t, \bm x)]\approx (\nabla\cdot\bm v[t, \bm x(t, \bm x+\bm r)]$ for $-t^*<t<0$ because the distance between the trajectories is much smaller than $l_0$. We set the values of divergences at the trajectory issuing at the midpoint of $\bm x$ and $\bm x+\bm r$ so as to have a symmetric form of the pair-correlation (the distinction between the points is beyond the accuracy of this calculation). Using that concentrations at distance $l'_0$ are not correlated (see below) we find,
\begin{eqnarray}&&\!\!\!\!\!\!\!\!
\langle n(\bm x)n(\bm x\!+\!\bm r)\rangle\!\approx \!\left\langle n[\!-t^*, \bm x(-t^*, \bm x)]\right\rangle \left\langle n[\!-t^*, \bm x(-t^*, \bm x\!+\!\bm r)]\right\rangle\nonumber\\&&\!\!\!\!\!\!\!\!\left\langle\exp\left[-2\int_{-t^*}^0 dt\nabla\cdot\bm v[t, \bm x(t, \bm x+\frac{\bm r}{2})]\right]\right\rangle,\!\label{paircorrelation}
\end{eqnarray}\\
see Fig. \ref{fig1}. Finally dividing the equation by its counterpart for $\langle n(\bm x))\rangle$,
\begin{eqnarray}&&\!\!\!\!\!\!\!\!
\langle n(\bm x)\rangle\!=\!\left\langle n[-t^*, \bm x(\!-t^*, \bm x)]\right\rangle\!\left\langle\!\exp\left[\!-\!\!\int_{-t^*}^0 \!\!\!\!dt\nabla\cdot\bm v[t, \bm x(t, \bm x)]\!\right]\!\right\rangle,\nonumber
\end{eqnarray}
where we can use $\bm x(t, \bm x+\bm r/2)$ instead of $\bm x(t, \bm x)$ we find,
\begin{multline}
f(\bm x, \bm r)\!\\
=\!\frac{\left\langle\!\exp\left[-2\!\int_{-t^*}^0\! dt \nabla\cdot\bm v[t, \bm x(t, \bm x\!+\!\frac{\bm r}{2})]\right]\right\rangle}
{\left\langle\exp\left[-\!\int_{-t^*}^0\! dt\nabla\cdot\bm v[t, \bm x(t, \bm x\!+\!\frac{\bm r}{2})]\right]\right\rangle^2}. \label{constr}
\end{multline}
This describes a positive correlation as accumulation of density increases due to motion in the same velocity divergence normalized by the accumulation that would occur due to motion in uncorrelated divergences. The latter determines $n_0(\bm x)$ but does not describe the "proper correlation" $f(\bm x, \bm r)$.

Since compressibility is small then we can find the averages in the RHS of Eq.\,(\ref{constr}) using the Gaussian averaging formula $\ln \langle \exp[x]\rangle=\langle x\rangle+\langle x^2\rangle_c/2$ where $\langle x^2\rangle_c=\langle x^2\rangle-\langle x\rangle^2$ is the dispersion (this neglects higher order cumulants of third order and higher in compressibility  \cite{Ma,fouxon1}). We find
\begin{eqnarray}&&\!\!\!\!\!\!\!\!\!\!\!\!
f(\bm x, \bm r)=\exp\left[t^*\int_{-\infty}^{\infty}\langle \nabla\cdot\bm v(0)\nabla\cdot\bm v(t)\rangle(\bm x) dt\right], \label{divvp1}
\end{eqnarray}
where we used that $t^*$ is much larger than the correlation time $\lambda^{-1}$ of $\nabla\cdot \bm v$, see Eq.~(\ref{tst}), and defined
\begin{eqnarray}&&
\langle \nabla\cdot\bm v(0)\nabla\cdot\bm v(t)\rangle(\bm x)=\lim_{t_0\to\infty}\frac{1}{t_0}\int_0^{t_0} dt' \nabla\cdot\bm v(t', \bm x)
\nonumber\\&&
\nabla\cdot\bm v[t'+t, \bm q(t'+t| t', \bm x)].
\end{eqnarray}
In the leading order in weak compressibility the definition uses the trajectories of the fluid (and not phoretic) particles that pass through $\bm x$ at time $t'$,
\begin{eqnarray}&&\!\!\!\!\!\!\!\!\!\!\!\!
\partial_t\bm q(t| t', \bm x)=\bm u[t', \bm q(t| t', \bm x)]\ \ \bm q(t=t'| t', \bm x)=\bm x,
\end{eqnarray}
cf. Eq.~(\ref{lagr0}). Using the definition (\ref{tst}) of $t^*$ in Eq.~(\ref{divvp1}) we find
\begin{eqnarray}&&\!\!\!\!\!\!\!\!\!\!\!\!
f(\bm x, \bm r)=\left(\frac{l'_0}{r}\right)^{\Delta(\bm x+\frac{\bm r}{2})}\approx \left(\frac{l_0}{r}\right)^{\Delta(\bm x+\frac{\bm r}{2})},
\end{eqnarray}\\
with $\Delta(\bm x)$ defined in Eq.~(\ref{inh}). Finally using that $\Delta\ll 1$, we obtain $(l'_0/l_0)^{\Delta}\approx 1$ where $l_0/l'_0\sim 10$ and thus finding Eq.~(\ref{parcor}). This formula holds when $r\ll l_0$, cf. the discussion around Eq.~(\ref{defintion}) and \cite{Batchelor}. In the case where the Batchelor scale is much smaller than the Kolmogorov one, the fluctuations of the concentration occur in much smaller regions of space than in the case of inertial particles. 

Similar considerations for higher-order correlation functions based on \cite{fouxon1} demonstrate that the log-normal statistics hold for rescaled concentration,
\begin{eqnarray}&& \langle {\tilde n}(\bm x_1){\tilde n}(\bm x_2)..{\tilde n}(\bm x_k)\rangle=\prod_{i>k}\langle {\tilde n}(\bm x_i){\tilde n}(\bm x_k)\rangle. \label{lognorm2} \end{eqnarray}

Furthermore the use of considerations of \cite{fouxon1} 
gives a deterministic solution and log-normal statistics for the coarse-grained concentration $n_l(0, \bm x)$ (cf. the next Section),
\begin{eqnarray}&& \frac{n_l(0, \bm x)}{\langle n(\bm x)\rangle}=\exp\left(-\int_{-t^*}^0 \nabla\cdot \bm v[t, \bm x(t, \bm x)]dt\right)\label{ct}\\&& \frac{\langle n_l^k(\bm x)\rangle}{\langle n(\bm x)\rangle^k}=\left(\frac{l_0}{l}\right)^{\Delta(\bm x)k(k-1)/2}, \label{log}\end{eqnarray}
where $n_l(t, \bm x)$ is defined with the help of the number of particles $N_l(t, \bm x)=\int_{|\bm x-\bm x'|<l}n(t, \bm x')d\bm x'$ inside the ball of radius $l\ll l_0$ centered at $\bm x$,
\begin{eqnarray}&& n_l(t, \bm x)=\frac{N_l(t, \bm x)}{(4\pi l^3)/3}, \label{eq:conc_def}\end{eqnarray}
so that for continuous distributions $l\to 0$ defines the concentration field (for the considered fractal distributions there is no well-defined limit). For $k=2$ Eq.~(\ref{log}) reproduces the scaling of the pair-correlation function because $\langle N_l^2(\bm x)\rangle=\int_{|\bm x-\bm x_1|<l,\ \ |\bm x-\bm x_2|<l}\langle n(t, \bm x_1)n(t, \bm x_2)d\bm x_1d\bm x_2\rangle$.

The derived pair-correlation implies that $\langle n(\bm x)n(\bm x+\bm r)\rangle$ has very different scales of variation with $\bm x$ and $\bm r$. The scale of variation with $\bm x$ is that of the average density profile which is determined by the scale $L$ of inhomogeneity of the statistics of turbulence. For spatially uniform statistics this dependence disappears. In contrast the dependence on $\bm r$ is a fast dependence that happens in the narrow range of $\bm r$ where the correlation function decays from infinite value at zero separation $\langle n^2(\bm x)\rangle=\infty$ to its large separation value $\langle n(\bm x)\rangle \langle n(\bm x+\bm r)\rangle$ at scales smaller than $l_0\ll L$ (there are no fluctuations at scale $l_0$ because of $\Delta\ll 1$).

Regularization of the divergence of $\langle n^2(\bm x)\rangle$ is determined by the breakdown of the continuity equation (Eq.\,\ref{continuity}) at the smallest scales. The breakdown can be determined by Brownian motion of the particles that introduces a diffusion term $D\nabla^2 n$ in the RHS of Eq.~(\ref{continuity}), by finite size of the particles, by the finite difference of the phoretic constants of the particles (due to size and properties difference) or other small scale phenomena.  Thus the divergent single-point dispersion $\langle n^2\rangle$ predicted by the power-law dependence is regularized at small scales at possibly large but finite value. The corresponding fluctuations of single-point concentration can be large with $\langle n^2(\bm x)\rangle$ larger than $\langle n(\bm x)\rangle^2$ by orders of magnitude.

\section{Pair correlations outside the scale of smoothness}\label{sec:five}

In this Section we consider the pair correlation function of concentration at all separations including those outside $l_0$. We use the consideration of \cite{fouxon1} that represents the steady state of concentration as the outcome of infinite time evolution starting with arbitrary initial condition where the concentration evolves according to the continuity equation. One starts with uniform initial condition $n(t=-T)=n_0$ in the remote past, finds $n(t=0)$ and takes the steady state limit of infinite evolution time $T\to\infty$. Solving the continuity equation along the particles' trajectories $\bm x(t, \bm x)$ defined in Eq.~(\ref{phortr}),
\begin{equation}
w(t, \bm x)=\nabla\cdot\bm v(t, \bm x)\nonumber \\
\end{equation}
\begin{multline}
\frac{d}{dt}n[t, \bm x(t, \bm x)]=[\partial_t+\bm v\cdot\nabla]n(t, \bm x)|_{\bm x=\bm x(t, \bm x)}=\\-n[t, \bm x(t, \bm x)]w[t, \bm x(t, \bm x)],
\end{multline}

we find,
\begin{eqnarray}&&\!\!\!\!\!\!\!\!\!\!\!\!\!\!
n(0, \bm x)=n_0\exp\left(-\int_{-T}^0 w[t, \bm x(t, \bm x)]dt\right).
\end{eqnarray}
We find for the pair-correlation function taking the product of $n(0, \bm x)$ and $n(0, \bm x+\bm r)$,
\begin{eqnarray}&&\!\!\!\!\!\!\!\!\!\!\!\!\!\! \frac{\langle n(\bm x)n(\bm x+\bm r)\rangle}{\langle n(\bm x)\rangle\langle n(\bm x+\bm r)\rangle}=\frac{1}{\left\langle\exp\left(-\int_{-T}^0 w[t, \bm x(t, \bm x)]dt\right)\right\rangle}\nonumber\\&&\!\!\!\!\!\!\!\!\!\!\!\!\!\!\frac{\left\langle\exp\left[-\int_{-T}^0 \left(w[t, \bm x(t, \bm x)]+w[t, \bm x(t, \bm x+\bm r)]\right)dt\right]\right\rangle}{\left\langle\exp\left(-\int_{-T}^0 w[t, \bm x(t, \bm x+\bm r)]dt \right)\right\rangle}.\end{eqnarray}
Using the cumulant expansion theorem for writing the averages we find that in the leading order in weak compressibility we can use the Gaussian approximation $\langle \exp[x]\rangle=\exp[\langle x\rangle+\langle x^2\rangle_c/2]$ in the averages \cite{Ma,fouxon1} which gives,
\begin{eqnarray}&&\!\!\!\!\!\!\!\!\!\!\!\!\!\! \frac{\langle n(\bm x)n(\bm x+\bm r)\rangle}{\langle n(\bm x)\rangle\langle n(\bm x+\bm r)\rangle}=\nonumber\\&&\!\!\!\!\!\!\!\!\!\!\!\!\!\!\exp\left[\int_{-\infty}^0\!\!dt_1dt_2 \langle w[t_1, \bm q(t_1, \bm x)]w[t_2, \bm q(t_2, \bm x+\bm r)]\rangle\right]\label{repr},\end{eqnarray}
where we took the steady state limit $T\to\infty$ and used the fluid particles trajectories $\bm q(t, \bm x)$ instead of $\bm x(t, \bm x)$ in the leading order in weak compressibility, see Eqs.~(\ref{lagr0})-(\ref{phortr}). This is a rigorous representation of the pair-correlation function in the limit of weak compressibility that was derived for spatially uniform statistics in \cite{fouxon1}. The pair correlation function at $r\ll l_0$ is obtained by observing that at these scales the flow divergence $w$ is identical at both trajectories up to times where the distance between the trajectories becomes comparable with the scale $l_0$ of spatial variations of $w(t, \bm x)$. If $r\to 0$ then the trajectories coincide at all times so we find divergence in $\langle n^2(\bm x)\rangle$. When $r$ is small but finite the leading order term is obtained considering how the time that the trajectories stay below $l_0$ diverges at small $r$. This is the time $t^*$ that we studied in the previous Section,
\begin{eqnarray}&&\!\!\!\!\!\!\!\!\!\!\!\!\!\! \int_{-\infty}^0\!\!dt_1dt_2 \langle w[t_1, \bm q(t_1, \bm x)]w[t_2, \bm q(t_2, \bm x+\bm r)]\rangle\approx\nonumber\\&&\!\!\!\!\!\!\!\!\!\!\!\!\!\! t^*\int_{-\infty}^{\infty} dt_2\langle w[t_1, \bm q(t_1, \bm x)]w[t_2, \bm q(t_2, \bm x)]\rangle,\ \ r\ll l_0, \label{ins}\end{eqnarray}
which reproduces formula (\ref{divvp1}) for the pair correlation function obtained in the previous Section (the pair correlation function in the integral on the RHS depends on time difference $t_2-t_1$ only because of incompressibility of flow of fluid particles). Beyond the scale $l'_0\sim l_0/10$ the spatial correlations of concentration are weak so that $\langle n(\bm x)n(\bm x+\bm r)\rangle\approx \langle n(\bm x)\rangle\langle n(\bm x+\bm r)\rangle$. This is reproduced by observing that at these scales the exponent in Eq.\,(\ref{repr}) is small because nothing compensates the smallness of compressibility (at smaller scales it is time of separation $t^*$ that does the compensation). The leading order correction is obtained using that $x-1\approx \ln x $ for $x\approx 1$ so,
\begin{eqnarray}&&\!\!\!\!\!\!\!\!\!\!\!
\frac{\langle n(\bm x)n(\bm x+\bm r)\rangle}{\langle n(\bm x)\rangle\langle n(\bm x+\bm r)\rangle}-1\approx \ln\!\left[\!\frac{\langle n(\bm x)n(\bm x\!+\!\bm r)\rangle}{\langle n(\bm x)\rangle\langle n(\bm x\!+\!\bm r)\rangle}\!\right] \nonumber \\&&\!\!\!\!\!\!\!\!\!\!\!\!\!\!=\int_{-\infty}^0 \langle w[t_1, \bm q(t_1, \bm x)]w[t_2, \bm q(t_2, \bm x+\bm r)]\rangle dt_1dt_2,\label{lead}\end{eqnarray} for $r\gtrsim l'_0$.
This formula holds for the Navier-Stokes turbulence involving no approximations so far. We provide estimates for the integral in the RHS.
We observe that at scales $r\gtrsim l_0$ the pair-correlation function $\langle w(\bm x)w(\bm x+\bm r)\rangle$ decays with $r$ in contrast with $r\ll l_0$. Thus if we introduce the characteristic time $t_r$ during which the separation of trajectories $\bm q(t_1, \bm x)$, $\bm q(t_2, \bm x+\bm r)$ grows by a factor of order one then we have,
\begin{eqnarray}&&\!\!\!\!\!\!\!\!\!\!\!\!\!\!
\frac{\langle n(\bm x)n(\bm x+\bm r)\rangle}{\langle n(\bm x)\rangle\langle n(\bm x+\bm r)\rangle}-1\sim t_r^2 \langle w(\bm x)w(\bm x+\bm r)\rangle \ r\gtrsim l'_0.
\end{eqnarray}
We disregarded the difference of time-scales of $l'_0$ and $l_0$ since it consists of logarithmic factor $\sim \ln 10$ which is of order one.
The scaling produced depends on $r/\eta$. If we consider $r\lesssim \eta$ than in time of order $\lambda^{-1}$ the trajectories separate by factor of order one so that we find,
\begin{eqnarray}&&\!\!\!\!\!\!\!\!\!\!\!\!\!\!
\frac{\langle n(\bm x)n(\bm x+\bm r)\rangle}{\langle n(\bm x)\rangle\langle n(\bm x+\bm r)\rangle}-1\sim \frac{\langle w(\bm x)w(\bm x+\bm r)\rangle}{\lambda^2} \ l'_0\lesssim r\lesssim \eta.\nonumber
\end{eqnarray}
When scales $r\gg \eta$ are studied the characteristic separation time $t_r$ of the trajectories separated initially (or rather finally) by $r$ roughly obeys the Richardson scaling $t_r\sim r^{2/3}\epsilon^{-1/3}$ so that
\begin{eqnarray}&&\!\!\!\!\!\!\!\!\!\!\!\!\!\!
\frac{\langle n(\bm x)n(\bm x+\bm r)\rangle}{\langle n(\bm x)\rangle\langle n(\bm x+\bm r)\rangle}-1\sim r^{4/3}\epsilon^{-2/3}\langle w(\bm x)w(\bm x+\bm r)\rangle,\label{pin}
\end{eqnarray}
where we disregard the corrections to Kolmogorov scaling that would become relevant at large Reynolds numbers. The self-consistency of the assumption of fast decay demands that $\langle w(\bm x)w(\bm x+\bm r)\rangle$ decays with $r$ faster than $r^{-4/3}$ - otherwise the pair-correlation function would not be a decaying function of the distance.

Formula (\ref{pin}) holds in all cases where the description with weakly compressible flow holds. If we use it in the case of inertial particles whose flow at small Stokes relaxation time $\tau$ (properly non-dimensionalized as small Stokes number) has $w=-\tau(\nabla_k u_i)(\nabla_i u_k)$ then the use of Kolmogorov scaling (dimensional analysis telling that the only time-scale of turbulence relevant at scale $r$ is $t_r$) gives $(\tau/t_r)^2$ for the RHS of Eq.~(\ref{pin}) that is  $\langle w(\bm x)w(\bm x+\bm r)\rangle\propto r^{-8/3}$ and 
\begin{eqnarray}&&\!\!\!\!\!\!\!\!\!
\frac{\langle n(\bm x)n(\bm x+\bm r)\rangle}{\langle n(\bm x)\rangle\langle n(\bm x+\bm r)\rangle}-1\sim \tau^2 \epsilon^{2/3}r^{-4/3}\label{prcr1}\\&& \!\!\!\!\!\!\!\!\!\langle n(\bm x)n(\bm x+\bm r)\rangle=\langle n(\bm x)\rangle\langle n(\bm x+\bm r)\rangle\exp\left[C\tau^2 \epsilon^{2/3}r^{-4/3}\right],\nonumber
\end{eqnarray}
where $C$ is a constant of order one, cf. Eqs.~(\ref{repr}),(\ref{lead}).
This scaling of the correlation function agrees with the prediction of the white noise model \cite{FFB} however here the result is obtained without modeling the flow so it holds for Navier-Stokes turbulence.

In the case of diffusiophoretic particles at large $\textrm Sc$ the range $l_d\ll\,r\ll\,\eta$ has no counterpart in the study of inertial particles. We have using Eq.~(\ref{lead}) with $w=D_p\nabla^2\ln C$,
\begin{eqnarray}&&\!\!\!\!\!\!\!\!\!
\frac{\langle n(\bm x)n(\bm x+\bm r)\rangle}{\langle n(\bm x)\rangle\langle n(\bm x+\bm r)\rangle}-1\approx D_p^2\int_{-\infty}^0\!\!  \langle \nabla^2\ln C[t_1, \bm q(t_1, \bm x)] \nonumber\\&&\!\!\!\!\!\!\!\!\!
\nabla^2\ln C[t_2, \bm q(t_2, \bm x\!+\!\bm r)]\rangle dt_1dt_2,\ \ l_d\ll r\ll \eta.\label{prcr}
\end{eqnarray}
We did not find a way for determining the $r-$dependence of the RHS. We can determine the order of magnitude using that at $r\sim l_0$ the correlation function has to agree with that at smaller scales given by Eq.~(\ref{ins}),
\begin{eqnarray}&&\!\!\!\!\!\!\!\!\!\!\!\!\!\!
\frac{\langle n(\bm x)n(\bm x+\bm r)\rangle}{\langle n(\bm x)\rangle\langle n(\bm x+\bm r)\rangle}-1\sim \Delta\left(\bm x+\frac{\bm r}{2}\right),\ \ r\sim l_0,\label{prcr2}
\end{eqnarray}
where the logarithmic factor in $t^*=|\lambda_3|^{-1}\ln(l_0/r)$ is of order one. The $r-$dependence of Eq.~(\ref{prcr}) could be determined using that the range of $l_d\ll r\ll \eta$ is characterized by one time-scale $\lambda^{-1}$ so that the integration times in the integral are of order $\lambda^{-1}$ giving,

\begin{equation}
\begin{aligned}
& \frac{\langle n(\bm x)n(\bm x\!+\!\bm r)\rangle}{\langle n(\bm x)\rangle\langle n(\bm x\!+\!\bm r)\rangle}\!-\!\!1\!\sim\! \\
& \frac{D_p^2}{\lambda^2}\langle \nabla^2\!\ln C(\bm x)\nabla^2\!\ln C(\bm x\!+\!\bm r)\rangle,
\end{aligned}
\label{eq:connect}
\end{equation}
that holds at $l_d\ll r\ll \eta$. Still, though the study of simultaneous correlation functions is simpler, we  could not determine the $r-$dependence of the correlation function in the last line. This includes trying to perform the study in the simplest context solving the equations on the correlation functions of $C$ obtained in the model where $C$ is considered as passive field with no reaction on the flow and the statistics of the flow is modeled as decorrelated in time but correlated in space. It is this model of statistics that helped finding solutions for other correlation functions but in this case significant obstacles obstruct the solution \cite{review}. However, we can use the obtained formula for experimental testing which will be done in Section \ref{section4}.

\section{Experimental confirmation of phoretic clustering}\label{section4}
According to the theory outlined above, phoresis leads to clustering of particles independent of the phoretic mechanism. These predictions generally relate to scales smaller than the turbulent smoothness scale $l_0$. Nevertheless effects are also expected above this scale which is the range investigated experimentally.
For the experimental analysis of phoretic clustering, we chose a turbulent flow with an inhomogeneous distribution of salinity in order to generate diffusiophoretic particle drift. This may also serve as a possible model for the formation of marine snow in the ocean.
In the experiment, we examine the existence and the degree of particle clustering by measuring the pair correlation function of particle concentration. Subsequently, we check the agreement of these experimental findings with the theoretical predictions provided in the previous Sections.

\subsection{Experimental technique}

\begin{figure*}
\centering
\includegraphics[width=0.78\textwidth]{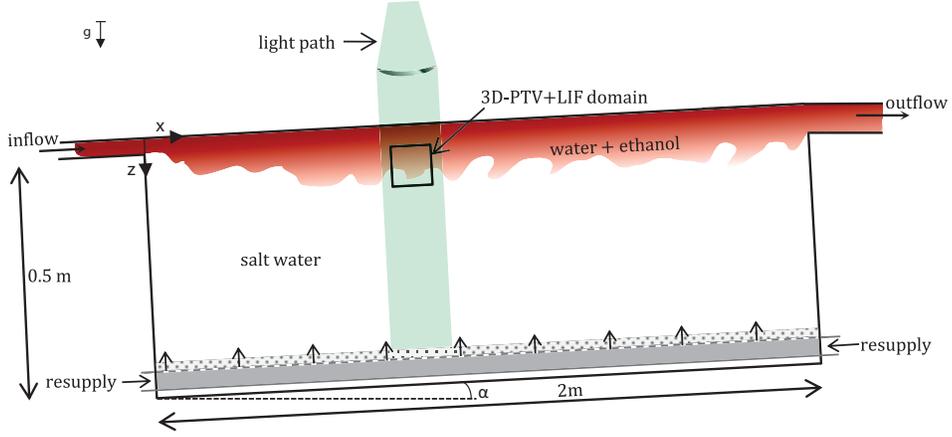}%
\caption{\label{fig:facility} Experimental setup showing the clear turbulent fluid (dark) rising along an inclined wall with angle $(\alpha = 10^\circ)$. Entrained salt water (white) is gently resupplied through two perforated pipes on the bottom of the tank.}
\end{figure*}

\begin{figure*}
\centering
\includegraphics[scale=0.43]{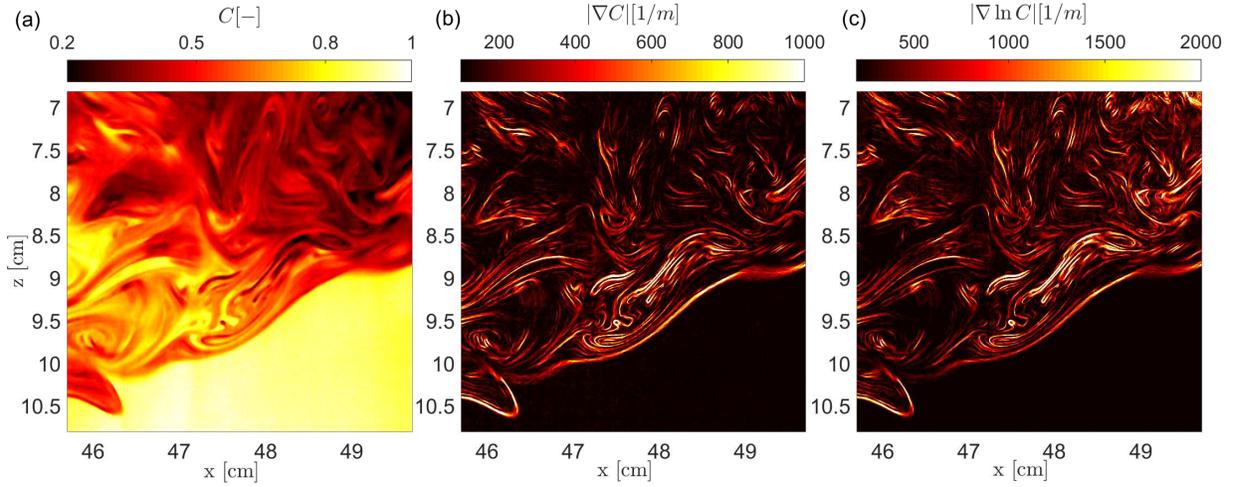}
\caption{\label{fig:contours} Instantaneous snapshots in the central plane of the observation volume of salt concentration field with $C=1$ being the maximum salt concentration (a), the corresponding norm of the gradient of the concentration field (b) and $|\nabla \ln C|$, which is the quantity that drives diffusiophoresis (c).}
\end{figure*}

An inclined gravity current setup is used to analyze the diffusiophoretic effect on particles in turbulent flow. The experimental setup is described in detail by Krug et al. \cite{krug2}. The facility, shown schematically in Fig.\,\ref{fig:facility}, allows for creation of a turbulent flow that features strong local gradients of salinity. The current is realized as a turbulent flow of water, mixed with 1.8 vol\% ethanol, rising along an inclined wall in a tank filled with dyed saltwater that is initially at rest. The small amount of ethanol in the light fluid serves to match the refractive indices of both fluids --- a crucial prerequisite for optical flow measurements. For simplicity we will nevertheless refer to the mixture of water and ethanol as ‘clear fluid’ in the following. An inhomogeneous salt distribution inside the turbulent fluid is created by entrainment of the saltwater into the lighter turbulent fluid from below. A representative snapshot of the resulting salt concentration field is presented in Fig.\,\ref{fig:contours}(a).

Employing a recently developed measurement technique \cite{krug1} that combines scanning 3D particle tracking velocimetry (PTV) and scanning laser induced fluorescent measurements (LIF) allows us to obtain both the velocity and the concentration along Lagrangian particle trajectories in 3D. The volume of investigation measures $4\times 2\times 4\,cm^3$ in the streamwise ($x$), the spanwise ($y$)  and the wall normal ($z$) direction, respectively. The PTV measurement is performed by recording particle images from four different viewing directions. Subsequently a stereoscopic matching of the recorded particles is done, which is then followed by  temporally connecting the obtained 3D particle positions. This provides the particle trajectories. The computation of the spatial and temporal velocity derivatives is based on a local linear interpolation of the velocity field and a weighted polynomial fit to the derivatives along particle trajectories \cite{luethi}. The linear interpolation relies on information from particles in close-by proximity of the investigated point. Therefore a sufficiently high particle seeding density is crucial to properly access the full Lagrangian velocity gradient tensor. A detailed description of the system was published by L{\"u}thi et al. \cite{luethi}.
Additionally, we measure the Laplacian of the salt concentration indirectly by using the total derivative along particle trajectories, i.e. using $dC/dt\,=\,D_s\nabla^2 C$. The spatial resolution of the velocity measurement is approximately $4\,\eta$ and the LIF resolution is $4\,l_d$ in the plane of the light sheet ($x\,-\,z$) and about $90\,l_d$ in the scanning ($y$) direction. These combined PTV and LIF measurements in three-dimensional space are used to analyze the evolution of $C$ along particle trajectories.

In order to characterize the flow field outside the limited spatial domain of the 3D measurements, simultaneous planar particle image velocimetry (PIV) \cite{krug2} and LIF measurements where performed on a domain located in the $x\,-\,z$ plane and the spanwise center of the tank. The domain extents $6\,cm$ in the wall- normal and streamwise direction respectively. This results in a pixel resolution of approximately $0.2\,\eta$. The particle seeding results on average in a inter-particle distance of $4 \eta$. One major advantage that these 2D techniques bring is the significantly longer recording time compared to the 3D measurements, where the scanning procedure limits the duration of the recording. This provides generally more statistics which is crucial for accurate measurements of the pair-correlation function of the particle concentration $n$. Therefore we used the planar PIV/LIF measurements for the analysis of the particle concentration $n$.

\begin{table}
\begin{ruledtabular}
\begin{tabular}{c c c}
\multicolumn{3}{|c|}{{\bf Properties of Flow and Particles}}\\
\hline\hline\\
$\textrm Re$ & $4800$ & $[-]$ \\
$\textrm Re_{\lambda_T}$ & $70$ & $[-]$ \\
$\lambda_T$ & $5.1\times10^{-3}$ & $[m]$\\
$\eta$ & $3\times10^{-4}$ & $[m]$\\
$\lambda$ & $11.1$ & $[1/s]$\\
$l_d$ & $10^{-5}$ & $[m]$ \\
$\mu$ & $10^{-3}$ & $[Pa \cdot s]$\\
$D_S$ & $1.99\times10^{-9}$ & $[m^2/s]$ \\
$\rho_p$ & 1016 & $[kg/m^3]$ \\
$\tau_p$ & $10^{-4}$ & $[s]$ \\
$d_p$ & $2\times10^{-5}$ & $[m]$\\
$D_p$ & $1.25\times10^{-9}$ & $[m^2/s]$ \\
$\textrm Sc$ & $500$ & $[-]$\\
$\textrm Stk$ & $10^{-3}$ & $[-]$\\
\hline
\end{tabular}
\end{ruledtabular}
\caption{Properties of the flow, the fluid, the salt and the particles used in the experiment. Where $\lambda_T$ is the Taylor microscale and $\textrm Re_{\lambda_T}$ the Taylor Reynolds number, respectively. The dynamic viscosity of the working fluid is written as $\mu$. The diffusivity of the salt is denoted as $D_S$. The important particle properties presented are the particle density $\rho_p$, the particle diameter $d_p$ and the resulting diffusiophoretic constant $D_p$, calculated according to \cite{andersonprieve}. The particle Stokes number $\textrm Stk$ is computed by the ratio of the particle response time $\tau_p$ to the Kolmogorov time scale $\lambda^{-1}$. The particle response time is defined as $\tau_p\,=\,\frac{2d_p^2\rho_p}{9\mu}$.}
\label{tab:flowprop}
\end{table}

The particles (VESTOSINT{\small\circledR} 2159 natural color, supplied by Evonik Industries AG) are uniformly seeded into the flow long before entering the gravity current test section. 
This guarantees a well mixed particle distribution.
The size of the particles used in the experiment is of $\mathcal{O}(l_d)$ and their $\zeta-$potential is approximately $-35mV$ (measured using a Zetasizer Nano Z), which implies a diffusiophoretic constant of $D_p\,\approx\,1.24\times10^{-9}m^2/s$ in NaCl according to \cite{andersonprieve}. Thus in the presence of salinity gradients, theory predicts that particles will acquire a drift velocity (see Table\,\ref{table}), which - according to our theoretical prediction - is expected to eventually induce clustering of particles. Relevant flow and particle properties of the experiment are summarized in Table\,\ref{tab:flowprop}.
Furthermore the particle concentration $n$ is a property of major importance to the following analysis. It is obtained experimentally by counting the number of particles in a circle with radius $2\eta$, according to the definition of Eq.\,(\ref{eq:conc_def}). This size of the circle is chosen in order to find a sufficient number of particles $(> 1)$ inside the circle.

An instantaneous snapshot of the normalized salt concentration $C$ is plotted in Fig.\,\ref{fig:contours}(a) where a value of 1 corresponds to unmixed salty water and thus the maximum salt concentration.
Snapshots of the norm of the corresponding instantaneous gradient of $C$ and $\ln C$ are shown in Fig.\,\ref{fig:contours}(b) \& (c), respectively. Since diffusiophoresis is driven by $|\nabla \ln C|$, the ubiquitous filaments of large values of $|\nabla \ln C|$ (in Fig.\,\ref{fig:contours}c) qualitatively indicate the regions where large diffusiophoretic velocities are expected. Note that the LIF measurements do not fully resolve the Batchelor scale and thus gradients are not sufficiently accurately resolved in this study. \\

\subsection{Profiles of average salt and particle concentration}\label{subsec:exp_profiles}

\begin{figure}[h]
\centering
\includegraphics[scale=0.18]{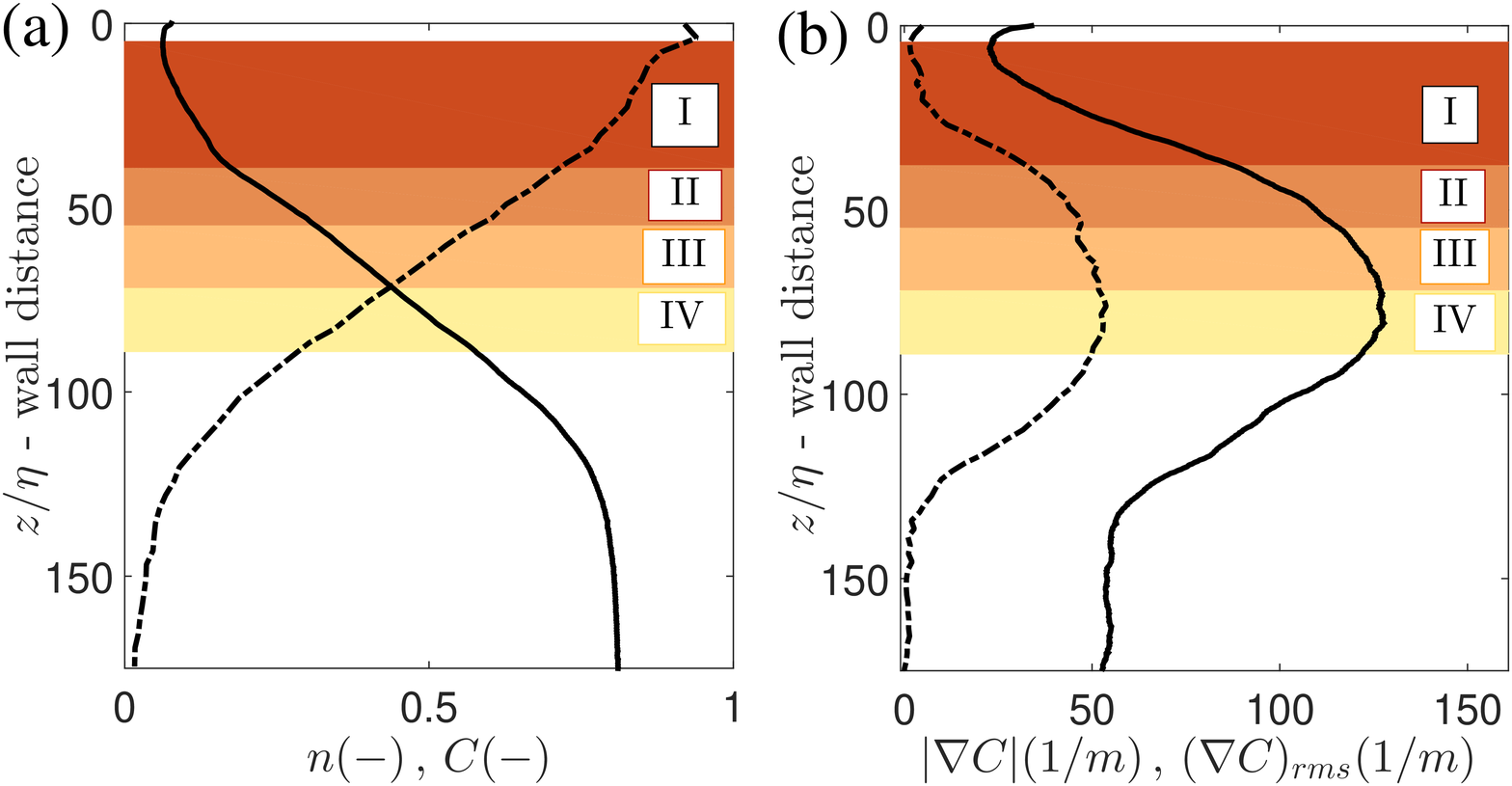}
\caption{Particle, salt concentration and salt concentration gradient averaged over time and homogeneous directions. (a) Averaged particle concentration $n(z)$ (dashed line) and salt concentration $C(z)$ (continuous line) - (b) averaged norm of the salinity gradient $|\nabla C(z)|$ (dashed line) and root-mean-square of the salinity gradient $(\nabla C(z))_{rms}$ (continuous line). The filled areas indicate 4 wall-parallel layers used for conditional averaging in Section \ref{subsec:exp_profiles}.\label{fig:partandsaltconc}}
\end{figure}

The gravity current has one inhomogeneous direction, namely the wall-normal $(z)$ direction, while $y$ is the homogeneous and $x$ the quasi-homogeneous coordinate. In particular, gradients of mean quantities are absent in $y-$direction and those in $x-$direction along the current have a characteristic length scale that is much larger than in $z-$direction. Profiles of particle and salt concentration averaged over the homogeneous directions are shown in Fig.\,\ref{fig:partandsaltconc}(a). Both, the average particle and the average salt concentration vary approximately linearly in depth over the so-called mixing layer ($\approx 40 \leq z/\eta \leq 100$). The average as well as the root-mean-square (rms) fluctuations of the salinity gradient shown in Fig.\,\ref{fig:partandsaltconc}(b) are also strongly dependent on $z$, where the former is significantly smaller than the latter. Due to the significant variations of the salinity gradient in wall-normal direction we expect the diffusiophoretic velocity and thus the degree of clustering to depend strongly on the $z-$direction. For a detailed analysis we divided our measurement domain in 4 regions according to the wall-normal distance. The regions are indicated by the colored areas in Fig.\,\ref{fig:partandsaltconc}, where dark colors generally represent lower salinity gradients and light colors high salinity gradients. We chose the regions such that $|\nabla C|$ is gradually increasing, as shown in Fig.\,\ref{fig:partandsaltconc}. Bin I (dark) corresponds to almost clear fluid, whereas bin IV (bright) contains the highest salt concentration and salt concentration gradients.

\subsection{Experimental results}

\begin{figure}[h]
\centering
\includegraphics[scale=0.55]{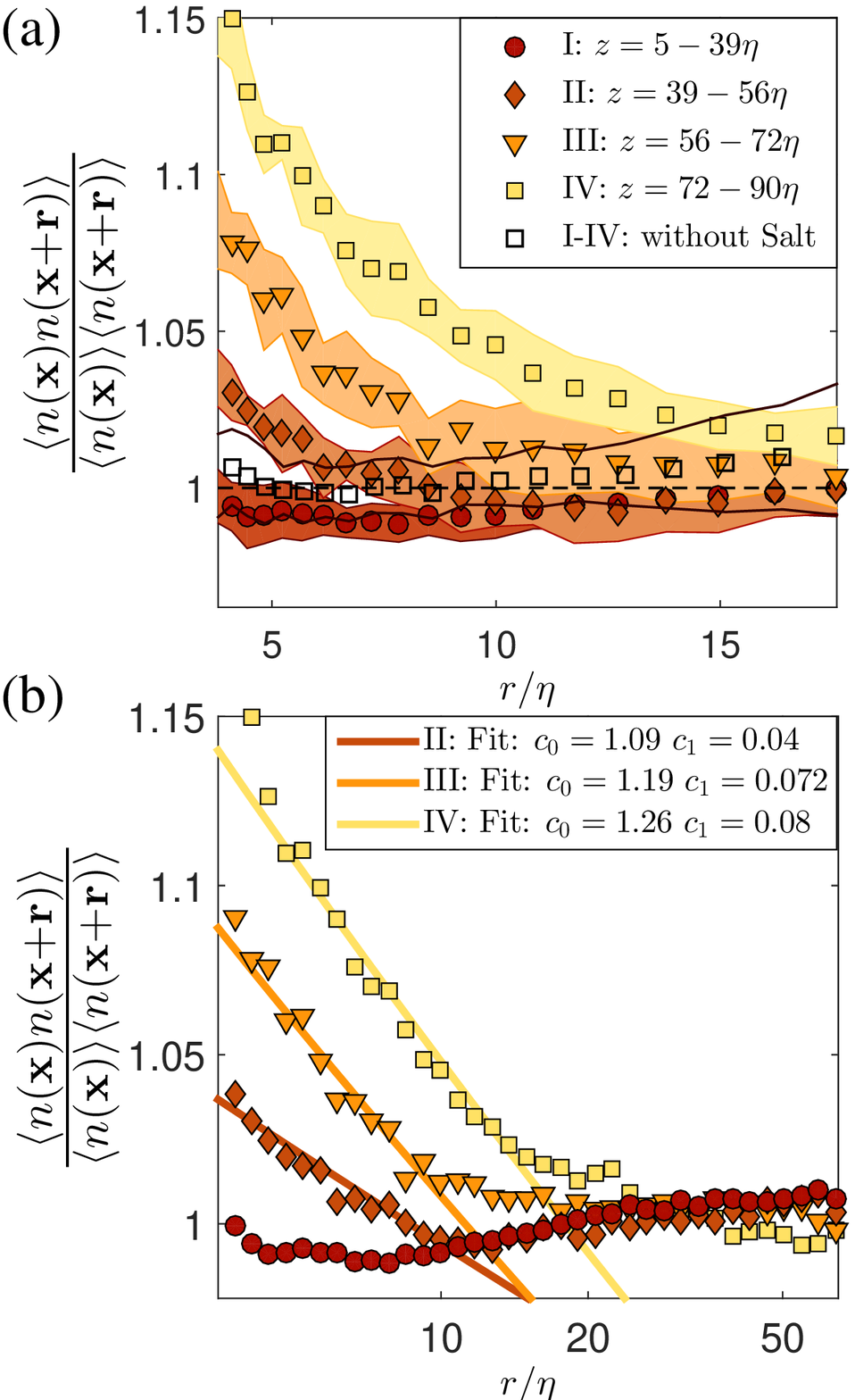}
\caption{\label{fig:2point_x1x2} Correlation of the particle concentration (Eq.\,\ref{log}), binned along $z$-direction in linear (a) and logarithmic (b) scale. The filled markers represent the pair-correlation in different regions in the gravity current flow, whereas the open, squared symbols are a spatially averaged pair correlation of the particle concentration in a flow without any salt. The shaded area in plot (a) indicates the standard deviation of this pair-correlation in streamwise direction. The colored, continuous lines in (b) indicate possible fits using $c_0\left(\frac{\eta}{r}\right)^{c_{1}(z)}$ for the different bins. The magnitude of the salinity gradient is generally low in bins marked with darker color and gradually increasing while going to lighter colors.} 
\end{figure}

In this Section we present the results of the experimental study on the pair-correlation function of the particle concentration. We obtain the pair-correlation of $n$ (particle concentration) from planar PIV/LIF measurements according to Eq.\,(\ref{eq:conc_def}), where we substitute $(4\pi l^3)/3$ by $\pi l^2$. The pair-correlation is evaluated in 4 regions that differ in wall-normal distance (to the upper wall of the facility) and correspondingly the magnitude of $\nabla C$. The results of this analysis are presented in Figure \,\ref{fig:2point_x1x2}. The pair correlation is plotted on a linear scale against the norm of the particle separation $\bm r$ in Fig.\,\ref{fig:2point_x1x2}(a) and on a semi-logarithmic scale in Fig.\,\ref{fig:2point_x1x2}(b). The shaded areas in the upper plot indicate the experimental uncertainty of these measurements. This uncertainty is quantified by the standard deviation of the pair correlation function along the quasi-homogeneous streamwise direction. The pair correlation in region I (circles) stays almost constant at approximately $1$ throughout all particle separations, indicating a lack of positive correlation caused by increase of the number of particles in regions with higher average concentration. That correlation is factored out considering correlations of normalized concentration $n(x)/\langle n(x) \rangle$. However, the pair-correlation of region II (diamonds), where significant concentration fluctuations are present, starts to deviate from $1$ at scales $r\leq8\,\eta$, showing minor inhomogeneities in the particle distribution at small separation scales. Further away from the top-wall, the data in regions III (triangular markers) and IV (filled squared markers) display a higher level of salinity gradients (Fig.\,\ref{fig:partandsaltconc}), which is accompanied by a stronger increase of the pair-correlation function at small particle separations in comparison to region II. The data points of region III and IV deviate from 1 at scales up to about $10\,-\,15 \eta$ (Fig.\,\ref{fig:2point_x1x2}). These results are a clear indicator for particle clustering with the clustering degree increasing as one moves to regions with larger distances to the upper wall of the experimental facility.

Since initially the particles are only seeded into the top part of the flow (without salt), it is important to check if the observed clustering might be caused by the mixing of the two fluids, that is to say, an inhomogeneous distribution of particles caused solely by turbulent transport of particles behaving as passive tracers. This would have nothing to do with diffusiophoresis in salinity gradients. To this end we conducted an experiment without adding salt, while keeping all other experimental conditions the same. Note that due to the active role of the salt this leads to a slightly different flow evolution. However the basic flow features, most notably the existence of a mixing shear layer, remain unchanged allowing us to gauge the potential impact of 'apparent clustering' due to mixing and the initial particle distribution. The corresponding data to these no-salt measurements is also included in Fig.\,\ref{fig:2point_x1x2}(a)(open squares). The continuous black lines in the same plot indicate the relevant measurement uncertainty of this data. The lack of clustering for the case of the spatially averaged curve of the flow without salt conclusively shows that mixing effects between the top turbulent fluid layer and the bottom non-turbulent fluid layer can be excluded as a cause of clustering in this experiment.

The pair correlations of the gravity current flow in the regions II-IV is well approximated by a power law of the form $c_0(\eta/r)^{c_1(z)}$. The fit parameters $c_0$ and $c_1$ for each bin are included in the legend of Fig.\,\ref{fig:2point_x1x2}(b). This power-law corresponds to the stretched exponential decay of the pair-correlation, cf. Eq.~(\ref{prcr1}). 
The exponent $c_{1}$ of the fit increases from $0$ in region 1 to $0.04$ in region 2 further to $0.072$ in region 3 up to $0.08$ in region 4. The pre-factor $c_0$ increases similarly for higher bin numbers. At first glance, these numbers appear rather small compared to what has been observed in various studies of inertial particle clustering (e.g. \cite{Collins,saw1,saw2}). However, comparing the diffusiophoretic velocity to the Kolmogorov velocity $v_{\eta}$ helps to classify these results properly. Considering that the maximum variation of $C$ from $0$ to $1$ physically occurs over Batchelor scale $(l_d\,=\,10^{-5}\,m)$, which results in values of order $10^{5}\,m^{-1} $ for $|\nabla C|/C$ (which is equivalent to $|\nabla \ln C|$) and thus a maximum diffusiophoretic velocity of $10^{-4}\,m/s$. The Kolmogorov velocity $v_{\eta}$ is computed as $\eta\times\lambda\,=\,3.33\times 10^{-3} m/s$. Defining a Stokes number $\textrm Stk$ for inertial particles as $\textrm Stk\,=\,\tau_p/\lambda^{-1}$, we similarly divide the maximum diffusiophoretic velocity by $v_{\eta}$, which results in a non-dimensional number that is $3\times 10^{-2}$. According to Saw et al. \cite{saw1}, inertial particles with $\textrm Stk$ similar to $3\times 10^{-2}$ have clustering exponents of order $10^{-2}$. This agrees well with the clustering exponents found in our study.
The results presented in Fig.\,\ref{fig:2point_x1x2} show that the degree of particle clustering strongly depends on the inhomogeneous coordinate as predicted by the theory. The observed increase of clustering with larger wall distance seems plausible in view of the fact that the salty water mixes into the less dense and non-salty fluid from below, leading to high concentration gradients at some distance from the top wall.

In order to support our conclusion that clustering arises solely due to diffusiophoresis it is useful to analyze the difference between inhomogeneities produced by turbulent transport of tracers and clustering because of preferential concentration. Turbulence can produce sharp contrasts and fronts in the tracers' concentration \cite{review}. However inside the contrasted regions there is no clustering. The pair-correlation function of tracers at decreasing distance between the points increases as a power-law but with an exponent whose sign is different from that observed in our experiment. For instance for the pair-correlation of tracer particle concentration $\theta$ (passive scalar) with spatially uniform statistics using the identity $2\langle\theta^2\rangle-2\langle\theta(0)\theta(r)\rangle=\langle[\theta(r)-\theta(0)]^2\rangle$ we find in the inertial range $\langle\theta(0)\theta(r)\rangle\,=\,\langle\theta^2\rangle-c\langle\theta^2\rangle(r/L)^{2/3}$ where $c$ is a constant of order one and $L$ is the scale at which scalar is injected.
Thus the pair-correlation function grows as power but with positive exponent, not the negative exponent that we observe in our experiments (Fig.\,\ref{fig:2point_x1x2}). It seems highly implausible that spatial non-uniformity of passive scalar would produce a power-law growth with negative exponent. This provides further confirmation that turbulent mixing and the initial seeding of particles are not responsible for the clustering we observe in our experiments.

In order to isolate diffusiophoresis as the only possible mechanism inducing clustering we have further analyzed the effect of the weak particles inertia ($\textrm Stk=10^-3$) using our PTV measurements. This three-dimensional measurements data allow us to compute the second invariant of the velocity gradient tensor $Q$, which is used to quantify the compressibility of the particle velocity that arises solely due to inertia. From the second invariant of the velocity gradient tensor we obtain the Laplacian of pressure $\nabla^2 p$ along particle trajectories. The integral of the temporal correlation of $\nabla^2 p$ along a particle trajectory is a measure for inertia-induced clustering \cite{fouxon1}. In our experiment this results in a scaling exponent $\Delta$ for inertial clustering on the order of $10^{-5}$. Hence, $\Delta$ due to inertial clustering is several orders lower compared to the scaling exponents we observe for the pair-correlation function ($c_1\,=\,\mathcal{O}(10^{-2})$) in Fig.\,\ref{fig:2point_x1x2}. 

We are therefore able to exclude inertia, mixing effects as well as the initial particle seeding as reasons for the experimentally observed clustering. We clearly note that the clustering degree increases significantly in regions with higher salinity gradients and did not appear at all in an experiment where the existence of diffusiophoresis is physically eliminated. Since diffusiophoresis is driven directly by gradients of the logarithm of the concentration gradient, we conclude that diffusiophoresis is the main driving force for the observed clustering.\\

We compare the observed pair-correlation function with the theory. The considered experimental situation is not exactly described by the theory since in the present experiment there is no scale separation between the size of the particles and the smallest (Batchelor) scale of turbulence. 
However the scales are of the same order of magnitude, thus the theory predicts that Eq.~(\ref{prcr}) holds by order of magnitude. We will further examine this in the following.

Substituting $\phi$ in the last formula of Eq.~(\ref{basic4}) with $\ln C$ leads to
\begin{eqnarray}&&\!\!\!\!\!\!\!\!\!\!\!\!\!\!
\partial_t\ln C+\bm u\cdot\nabla \ln C=D_S\nabla^2 \ln C+D_S\left(\nabla \ln C\right)^2. \label{drln}
\end{eqnarray}
We make the assumption that both terms on the RHS of Eq.\,(\ref{drln}) have identical scaling in the correlation function so that the $r-$dependence of the pair-correlation function in Eq.~(\ref{prcr}) at $l_d\ll r\ll \eta$ can be obtained using,
\begin{eqnarray}&&\!\!\!\!\!\!\!\!\!
\frac{\langle n(\bm x)n(\bm x+\bm r)\rangle}{\langle n(\bm x)\rangle\langle n(\bm x+\bm r)\rangle}-1\propto D_p^2\int_{-\infty}^0\!\!
\left\langle \left[\nabla^2 \ln C+\left(\nabla \ln C\right)^2\right] \right.\nonumber\\&&\!\!\!\!\!\!\!\!\!
[t_1, \bm q(t_1, \bm x)]\left[\nabla^2 \ln C+\left(\nabla \ln C\right)^2\right][t_2, \bm q(t_2, \bm x\!+\!\bm r)]\rangle dt_1dt_2,\nonumber
\end{eqnarray}
where proportionality designates that both sides have similar $\bm r$ dependence. This assumption is highly plausible because the ratio of $\nabla^2 \ln C$ and $\left(\nabla \ln C\right)^2$ involves the slowly varying field $\ln C(\bm x)$ that changes over the scale $\eta$. Then using the material derivative along the fluid particle trajectory yields,
\begin{eqnarray}&&
\frac{d}{dt}\ln C[t, \bm q(t, \bm x)]=\left[\partial_t C+\bm u\cdot\nabla\right] \ln C,
\end{eqnarray}
with Eq.~(\ref{drln}) we find,
\begin{eqnarray}&&\!\!\!\!\!\!\!\!\!
\frac{\langle n(\bm x)n(\bm x+\bm r)\rangle}{\langle n(\bm x)\rangle\langle n(\bm x+\bm r)\rangle}-1\sim \frac{D_p^2}{D_S^2}\lim_{\Delta t\to\infty}\int_{-\Delta t}^0 dt_1dt_2  \nonumber\\&&\!\!\!\!\!\!\!\!\!\left
\langle \frac{d}{dt_1}\ln C[t_1, \bm q(t_1, \bm x)]
\frac{d}{dt_2}\ln C[t_2, \bm q(t_2, \bm x\!+\!\bm r)]\right\rangle \nonumber\\&&\!\!\!\!\!\!\!\!\!
=\frac{D_p^2}{D_S^2}\lim_{\Delta t\to\infty}\left\langle \ln\left(\frac{C(0)}{C(-\Delta t)}\right)_{1}\ln\left(\frac{C(0)}{C(-\Delta t)}\right)_{2}\right\rangle,
\label{eq:correlation_lnC}
\end{eqnarray}
where the indices $1$ and $2$ denote the two particles' trajectories that pass through the points $\bm x$ and $\bm x+\bm r$ at $t=0$. The correlation function in Eq.~(\ref{eq:correlation_lnC}) is symmetric with respect to the change $\Delta t\to -\Delta t$ due to the incompressibility of turbulence.

We remark that the integral of the correlation function of time derivatives $d\ln C/dt$ vanishes when $r=0$ because it becomes the integral of a complete time derivative. However this degeneracy disappears in the range $r\gtrsim \eta$ (as confirmed by experimental observations) so that the order of magnitude estimates above, hold.

It is this theoretical prediction (\ref{eq:correlation_lnC}) that we use for comparison with the experiment. This form uses solely the actual concentration field and does not involve the spatial differentiation of the salinity field. This differentiated field would increase the error significantly since $|\nabla C|$ varies over the smallest, (Batchelor) scale $l_d$, which is not fully resolved in our measurements. The effect of this under-resolution on the salt concentration field itself which we use for our computation, results in a deviation of the variance of $C$, $\sigma^2(C)$, of less than $0.5\%$ compared to the fully resolved field.  Compared to other sources of error in our measurement such as pixel noise, particle attenuation and stripes (discussed in detail in \cite{krug2}) the filtering effect of the salinity field is significantly lower. These measurements therefore allow us to perform qualitative estimations of the clustering and compare those with the pair-correlation results shown in Fig.\,\ref{fig:2point_x1x2}.

To evaluate Eq.\,(\ref{eq:correlation_lnC}) experimentally, we considered the salt concentration along pairs of trajectories separated by a distance $\bm r$ at a specific time $t$. Therefore, we find all particle pairs separated by a certain distance $r\pm 0.1\,r$ at any time $t$ and track this trajectory pair for a time interval $\Delta t$ from this time step $t$ on. This analysis requires the evaluation of 3D data. Therefore the results of the combined scanning PTV/LIF are used for the following computation. We determined the salt concentration along the trajectories at times $t$ and $t\pm \Delta t$, subsequently taking the average over all pairs of trajectories. To compensate for the fact that particle movement is determined by the diffusiophoretic particle constant $D_p$ and not the measured salt diffusion $D_S$, we multiply the correlation by the factor $D_p^2/D_S^2$, which is $0.395$.

\begin{figure}
\centering
\includegraphics[scale=0.42]{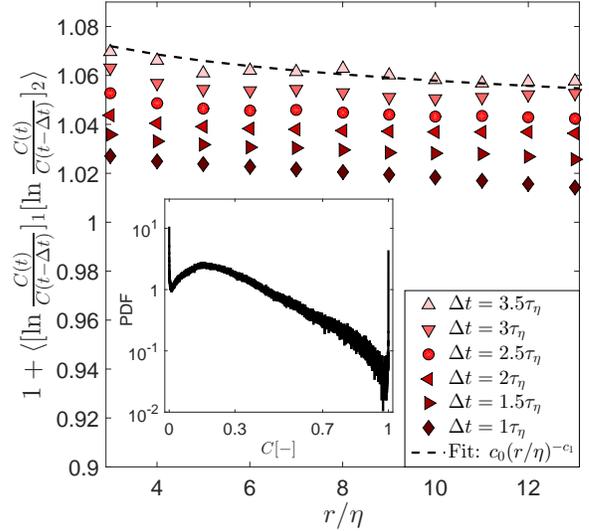}
\caption{\label{fig:divlnC}  Spatial correlation of $\ln C$ along pairs of trajectories plotted against the particle separation distance. Increasing $\Delta t$ increases also the absolute value of the correlation. The power-law behavior of the data at $\Delta t = 3.5 \tau_\eta$ is approximated by the dashed line, with the coefficients $c_0\,=\,1.085$ and $c_1\,=\,0.011$.
Inset: PDF of $C$ showing fluctuations over two orders of magnitude.}
\end{figure}

The correlation of Eq.\,(\ref{eq:correlation_lnC}) as a function of the particle separation $\bm r$ is shown in Fig.\,\ref{fig:divlnC}. We did not study the dependence of $\langle n(\bm x)n(\bm x+\bm r)\rangle$ on the direction of $\bm r$ because of the strong anisotropy of the flow (difference between vertical and horizontal directions). This would demand a significantly larger pool of data and is beyond the scope of the present work. The different markers in Fig.\,\ref{fig:divlnC} represent different $\Delta t$ over which the salinity concentration has been tracked along pairs of trajectories. The curves shift upwards with increasing $\Delta t$. The magnitude of all curves decreases slowly at larger particle separation $r$. The robustness of this calculation strongly depends on the number of pairs that are tracked. This is caused by the strong fluctuations of $C$, which are captured in the PDF of $C$ that is presented as an inset in Fig.\,\ref{fig:divlnC}. We find that sufficient robustness of the curves is guaranteed only by tracking a minimum of $2000$ pairs of trajectories for each step in time and space. All the data presented in Fig.\,\ref{fig:divlnC} fulfills this criterion. The number of trajectory pairs analyzed increases significantly with lower $\Delta t$. For most of the particle separations of the curves within $1\,\tau_{\eta}\leq \Delta t \leq 2.5\,\tau_{\eta}$ we find more than $10^{4}$ pairs of trajectories. We obtained good convergence for the limit of infinite $\Delta t$ over few Kolmogorov time-scales $\lambda^{-1}$, cf. Eq.\,(\ref{eq:connect}) and Fig.\,\ref{fig:divlnC} discussed below. The full convergence of $\Delta t$ independent of the limit is expected to occur at about $5 \tau_{\eta}$ \cite{fouxon1} which is out of reach for the present experiment due to the limited length of trajectories. This explains why the curves at different $\Delta t$ do still show a slight increase as one goes to longer tracking times. This vertical shift is expected to vanish as the tracking time approaches $\Delta t\,=\,5\eta$. The statistical difficulties are also the reason why we did not take into account any dependence on the wall-normal direction here. However, our analysis showed that also here the trajectories in regions of high salinity gradients are generally shifted upwards compared to trajectories in lower salinity regions. Due to poor statistical convergence we do not present these results here.

The dashed line in Fig.\ref{fig:divlnC} is a power law fit of the form $c_0(\eta/r)^{c_1}$ to the curve at $\Delta t= 3.5\tau_\eta$ with $c_1=0.011$ and a prefactor $c_0$ of $1.085$. In order to confirm the order of magnitude of the results obtained from the pair-correlation function we compare the prefactor $c_0$ obtained for the measurements of Fig.\,\ref{fig:divlnC} with the prefactors that were found for the different bins of the pair-correlation function (where we obtained values between $1$ and $1.26$ for $c_0$ depending on the $z-$position). Since the results in Fig.\,\ref{fig:divlnC} represent an average over the wall-normal distance we conclude that the order of magnitude estimates hold and the experiment qualitatively (and by the order of magnitude quantitatively) confirms the theoretical prediction (\ref{eq:correlation_lnC}).

Finally we compare the order of magnitude of the pair-correlation function with $\Delta$ which is predicted to give the order of magnitude of the pair correlation at the scale $l_d$, see Eq.~(\ref{prcr2}). We have
\begin{multline}
\Delta(\bm x)=\frac{D_p^2}{|\lambda_3|}\int_{-\infty}^{\infty}\!\!  \langle \nabla^2\ln C(0)\nabla^2\ln C(t)\rangle dt \\ \sim \frac{D_p^2}{\lambda^2}\langle |\nabla^2\ln C|\rangle^2 \nonumber
\end{multline}
and we use this to approximate $\Delta$ by
\begin{eqnarray}&&
\Delta(\bm x)\sim  \frac{D_p^2}{\lambda^2D_S^2} \left\langle \left|\frac{d\ln C}{dt}\right| \right\rangle^2.
\end{eqnarray}
Using the averaged material derivative along all trajectories, we find $(D_p^2/\lambda^2 D_S^2)\langle |d(\ln C)/dt| \rangle^2$ in our experiment. Similar to our previous observations these results depend strongly on the inhomogeneous coordinate and we observe significantly larger values for $\Delta$ in regions with strong salinity gradients compared to the low-gradient regions. For further comparison with the previous results we average $\Delta$ over the inhomogeneous flow direction and obtain that $\Delta$ is approximately $0.15$. Comparing this value with the spatially averaged value of the prefactor $c_0-1$ of the measurements of the pair-correlation function we conclude that
\begin{eqnarray}&&\!\!\!\!\!\!\!\!\!
\frac{\langle n(\bm x)n(\bm x+\bm r)\rangle}{\langle n(\bm x)\rangle\langle n(\bm x+\bm r)\rangle}-1\sim \Delta \ r\sim \eta.
\end{eqnarray}
This is true because $c_0-1$ from Fig.\,\ref{fig:2point_x1x2}(b) averaged over all bins and weighted by their size results in $\langle c_0-1 \rangle\,=\,0.11$. Note that these numbers are not definite numbers. Our approach should rather be seen as a way of observing the effect qualitatively and connecting theory and experiments throughout the whole range of scales.
Comparison with Eq.~(\ref{prcr2}) leads to the conclusion that the pair-correlation function does not decrease by order of magnitude in the Batchelor range $l_d<r<\eta$. This is quite reasonable since $\eta/l_d\sim 10$ is not too large.

\section{Discussion and Conclusions} \label{sec:conclusions}

In this paper we studied the behavior of diffusiophoretic particles in turbulent flow. The regime of fast reaction where the particles follow the local flow up to the diffusiophoretic drift has been investigated intensively. We demonstrated that the theory of clustering in weakly compressible flows applies. This implies fractality of the particles' distribution in space (clustering). The provided theory does not include the practically relevant case where the particles' size is comparable with the smallest spatial scale of turbulence (the Batchelor scale in the case that we studied). Thus we performed experiments in that range, confirming that the clustering continues holding though the theory works only by order of magnitude.

We demonstrated that phoretic particles that perform steady motion at a constant velocity $v_{ph}\,=\,c_{ph}\nabla\phi$ in the presence of gradients of the field $\phi$ in a fluid at rest, will move in the flow $\bm u(t, \bm x)$ with the speed
\begin{eqnarray}&&
\bm v(t)=\bm u[t, \bm x(t)]+c\nabla \phi[t, \bm x(t)],\label{flw}
\end{eqnarray}
provided that the characteristic temporal and spatial scales of the flow are above the characteristic scales of the phoretic phenomenon. The particles' motion in space fits the frame of weakly compressible flow,
\begin{eqnarray}&&
\frac{d\bm x}{dt}=\bm v[t, \bm x(t)],\ \ |\nabla\cdot \bm v|\ll |\nabla \bm v|.
\end{eqnarray}
Thus we could use the theory of the distribution of particles in weakly compressible random flows \cite{fouxon1,FFS} for predicting that phoretic particles distribute in space over a multi-fractal set with the pair-correlation function of the particle concentration $n$ obeying
\begin{eqnarray}&&\!\!\!\!\!\!\frac{\langle n(\bm x)n(\bm x+\bm r)\rangle}{\langle n(\bm x)\rangle\langle n(\bm x+\bm r)\rangle}=\left(\frac{l_0}{r}\right)^{\Delta(\bm x+(1/2)\bm r)} \ r\ll l_0,\label{inh1} \end{eqnarray} where $l_0$ is the smallest of the Kolmogorov and Batchelor scales. The function $\Delta(\bm x)$, given by Eq.~(\ref{inh}), varies in space over the scales of inhomogeneity of the statistics of turbulence. The positivity of $\Delta(\bm x)$ signifies divergence of the rms fluctuation of the concentration ($\langle n^2(\bm x)\rangle=\infty$) which manifests the distribution of particles over a singular multifractal set in space.

In the range $r\gtrsim l_0$ the correlations are weak, $\langle n(\bm x)n(\bm x+\bm r)\rangle\approx \langle n(\bm x)\rangle\langle n(\bm x+\bm r)\rangle$. We determined the correction that corresponds to (where it can be proved) stretched exponential decay of $\langle n(\bm x)n(\bm x+\bm r)\rangle/ \langle n(\bm x)\rangle\langle n(\bm x+\bm r)\rangle$ to $1$.

Equation~(\ref{inh1}) goes beyond the previous theory developed for spatially uniform statistics of turbulence \cite{fouxon1}. It holds in the case of inhomogeneous turbulence provided the characteristic scale of inhomogeneity is not too small. Thus it can be used in a wide range of situations, including the experiment performed in this work.


Our theoretical predictions apply to a wide range of phoretic phenomena. These include thermophoresis, diffusiophoresis, chemotaxis and electrophoresis. The prediction is based on the framework of weakly compressible flow that already proved itself in the study of intertial particles and phytoplankton \cite{Nature2013,fl}. However, the study of inhomogeneous turbulence and extension outside the scale of smoothness, crucially extending the validity of the theory to many practical applications, is the content of this work only.

Using simultaneous 3D particle tracking and concentration measurements in a turbulent gravity current, we confirmed the theoretical prediction of phoretic clustering in turbulent flow. We measured positive pair correlations of diffusiophoretic particles that increase in regions of higher salinity gradients. Further, we confirmed these measurements using correlations of concentration over pairs of trajectories as well as order of magnitude estimates for gradients of concentration. Given that the particle size used in the experiment is of the same order as the Batchelor scale $l_d$, the experiments demonstrate clustering beyond the theoretical limits which increases the practical relevance and range of applications. 

The observed diffusiophoretic clustering could have an effect on the formation of marine snow. In the ocean $l_d$ can well be of the order of the size of colloidal particles or smaller. The typical value for the energy dissipation $\epsilon$ per unit volume per unit time in oceanic flows is $10^{-6}m^2/s^3$. Therefore the Kolmogorov time-scale is $\tau_{\eta}=\sqrt{\nu/\epsilon}\sim 1 s$. Correspondingly $\eta\,=\,\nu^{3/4}\epsilon^{-1/4}\,=\,\tau_{\eta}^{3/2}\sqrt{\epsilon}\sim 10^{-3}m$, which gives $l_d\sim 10^{-5} m$. This length scale is comparable to typical sizes of colloidal particles in the ocean \cite{Jackson}. We therefore suggest that diffusiophoresis may accelerate the agglomeration of organic matter and formation of marine snow. \\

Financial support from the Swiss National Science Foundation (SNSF) under Grant No. 144645 is gratefully acknowledged. We thank Thomas Ki{\o}rboe for helpful discussions.

\bibliography{biblio}


\end{document}